\documentclass[useAMS,usenatbib]{mn2e}
\usepackage{txfonts}
\usepackage{graphicx}
\usepackage{natbib}
\usepackage{if then}

\def\del#1{{}}

\newcommand{\unit}[1]{\mbox{  }\rm{#1}}

\newcommand{\vect}[1]{{\vec {#1} }}

\title{Digital Signal Processing in Cosmology}
\author[Jens Jasche, Francisco S. Kitaura , Torsten A. En\ss lin]
       {Jens Jasche $^{1}$, Francisco S. Kitaura $^{1,2}$,Torsten A. En\ss lin $^{1}$ \\$^{1}$ Max-Planck-Institut f\"{u}r Astrophysik , Karl-Schwarzschild Strasse 1,  D-85748 Garching, Germany\\$^{2}$ Scuola Internazionale Superiore di Studi Avanzati , via Beirut 2-4,  I-34014 Trieste, Italy\\}
\begin{document}
\date{Submitted to MNRAS 20-Jan-2009}

\pagerange{\pageref{firstpage}--\pageref{lastpage}} \pubyear{2006}

\maketitle

\label{firstpage}

\begin{abstract}
We address the problem of discretizing continuous cosmological signals such as a galaxy distribution for further processing with Fast Fourier techniques. Discretizing, in particular representing continuous signals by discrete sets of sample points, introduces an enormous loss of information, which has to be understood in detail if one wants to make inference from the discretely sampled signal towards actual natural physical quantities. We therefore review the mathematics of discretizing signals and the application of Fast Fourier Transforms to demonstrate how the interpretation of the processed data can be affected by these procedures. 
It is also a well known fact that any practical sampling method introduces sampling artifacts and false information in the form of aliasing. 
These sampling artifacts, especially aliasing, make further processing of the sampled signal difficult.
For this reason we introduce a fast and efficient supersampling method, frequently applied in 3D computer graphics, to cosmological applications such as matter power spectrum estimation. This method consists of two filtering steps which allow for a much better approximation of the ideal sampling procedure, while at the same time being computationally very efficient.Thus, it provides discretely sampled signals which are greately cleaned from aliasing contributions.
\end{abstract}


\section{Introduction}
Many cosmological applications, like estimating the power spectrum from galaxy surveys or calculating the gravitational potential of the dark matter distribution in numerical simulations, rely on the use of Fast Fourier Transform (FFT) techniques. They are favored above other computational techniques as their low computational complexity allows to process huge datasets in reasonable times.

However, since by definition the FFT operates only on finite sets of discrete sample points the problem arises how to discretize and digitize a continuous signal, in order to record it in computer memory and make it available for further processing with FFTs. Natural physical signals, though, are generally neither discrete in real-space nor in Fourier-space.

Reducing such a signal to a set of finite and discrete sample points will naturally introduce an enormous loss of information. For this reason we try to answer what kind and how much information of the true physical system is still represented by the sampled data. This is equivalent to ask for a response function describing the operation of the computer data acquisition.

According to Shannon's theorem sampling a continuous signal can be achieved by low-pass filtering and sampling the continuous signal at discrete positions \citep{Shannon48,Shannon49}.
Low-pass filtering requires to convolve in real-space with the sinus cardinal function, which is infinite in extend, and is therefore not practical for real world applications. For this reason one usually relies on filter approximations of the ideal low-pass filter, which only have finite support in real-space and hence allow for fast computation of the low-pass filtering convolution.

In cosmology some frequently used techniques to discretize a continuous distribution of point particles are Nearest Grid Point (NGP), Cloud In Cell (CIC) or Triangular Shaped Clouds (TSC) \citep{HOCKNEYEASTWOOD1988}, which all attribute a weighted fraction of the individual point particle mass to the surrounding discrete grid positions.
The process of sampling by relying on these filter approximation in general not only irrevocably reduces the information content of the continuous signal to that one represented by a set of discrete points, it also introduces sampling artifacts like aliasing or Gibbs ringing \citep{Wolberg1997,Marschner1994,Volker2008}.
It is often the leakage of aliasing power from the stop-band which makes further processing of the sampled signal difficult. For example, estimation of higher-order spectra, like the bi- or tri-spectrum, with FFT techniques will be erroneous due to sampling artifacts. As mentioned in \citet{Volker2008}, so far, there is no known approach to accurately correct for these sampling effects in measuring higher-order spectra like the bi-spectrum with FFT techniques.
The aim for signal processing technologies therefore is to find low-pass filter approximations which sufficiently suppress these sampling artifacts, while at the same time still being computationally far less expensive than the ideal sampling procedure. The Digital Signal Processing (DSP) literature provides plenty of possible approaches \citep[see e.g.][]{DSPGUIDE2002} which could be introduced to various cosmological signal processing problems.

For cosmological applications recently different approaches have been presented to deal with the problem of sampling artifacts \citep{Jing2005,Volker2008}. \citet{Jing2005} proposes to evaluate all aliasing sums of the  NGP, CIC and TSC method. He finds that the sampled shot noise contribution to the power spectrum can be expressed in a simple analytic way. To correct for further aliasing contributions, he introduces an iterative correction method for the power spectrum, assuming a power-law behavior. However, Jing's method is only applicable to power spectrum estimations and in cases where there are no other sources of mode-coupling, like masks of a galaxy survey or selection functions. The method proposed by \citet{Volker2008} utilizes the scale functions of the Daubechechies wavelet transformation as filter approximations. It has been demonstrated to yield better results than standard CIC or TSC methods.

However, most of these approaches rely on increasing the spatial support of the filter approximation to more closely represent the properties of the ideal low-pass filter. This approach is well known in the DSP literature, which favors the windowed sinc functions as optimal approximations to the ideal low pass filter \citep{Hauser2000,Duan2003,Marschner1994,DSPGUIDE2002}.
These approaches, though, become computationally more expensive as the spatial support of the filter approximation grows, and tends towards evaluating the ideal low-pass filtering convolution.
Many applications, like for instance particle mesh codes, however, rely on fast but nevertheless accurate low-pass filtering procedures. Hence, sampling a continuous signal like the galaxy distribution on a discrete grid to further process it via FFT techniques is always a trade off between computational efficiency and quality.

For this reason we introduce a new supersampling technique frequently applied as an anti-aliasing technique in 3D computer graphics to cosmological applications like power spectrum estimation \citep{Wolberg1997,GOSS2000}.

Our method is a two step filtering process in which the signal first is sampled to a discrete grid with super resolution via the CIC or TSC method, then low-pass filtered with the ideal discrete low-pass filter, and finally resampled at target resolution.

In this fashion, by using simple low-pass filter approximation and utilizing FFTs, we provide a fast method to accurately calculate a low-pass filtering procedure. 

As our method uses existing finite support filter approximation, it can be understood as complementary to the approach of finding better low-pass filter approximations. The combination of both approaches will naturally lead to more optimal results.
However, the supersampling method in itself is computationally much less expensive than increasing the spatial support of the filter approximation, while at the same time presenting better results as can be demonstrated in the case of cosmological matter power spectrum estimation.

Beside being a numerically efficient method, the supersampling procedure does not only aim at correcting aliasing effects in the power spectrum, it also greately cleans the discretized signal from aliasing contributions.

This is important for applications which operate on the sampled signal, like calculating the gravitational potential for particle mesh simulations \citep{HOCKNEYEASTWOOD1988}, or for further processing the discrete galaxy distribution within a FFT based linear Wiener filtering process  as described in \citet{Kitaura}.

This paper is organized as follows.
We begin by briefly outlining the general requirements for applying FFTs to real continuous signals in section \ref{The_requirements_of_FFTs}. A review of Shannon's sampling theorem and the discussion of how to discretize signals sampled to a bounded domain are given in section \ref{Descretize_Rspace}. In section \ref{DISCRETIZING_FOURIERSPACE} we discuss the Fourier space discretization of signals, and show, that while it is not possible to uniquely recover the entire signal from the sampled dataset, it is possible to uniquely restore individual Fourier modes, when applying the appropriate filter method. We also show that the discretized signal in general will not reflect all physical properties of the underlying continuous signal. In section \ref{Sampling_3d_galaxy_distributions} we outline the ideal sampling procedure, while in section \ref{practical_sampling} we discuss the actual problems of practical sampling procedures. In section \ref{Supersampling} we describe our supersampling method, and show how this method can help to aliviate aliasing contributions.
This is followed by a discussion in section \ref{Discussion}.

\begin{figure*}
	\centering
	{
	\begin{picture}(100,240)

\put(-180,240){a)}

  	\put(-180,240){
	\rotatebox{270}
	{
		\includegraphics[bb = 61 170 552 650,width=0.2\textwidth,clip=true]{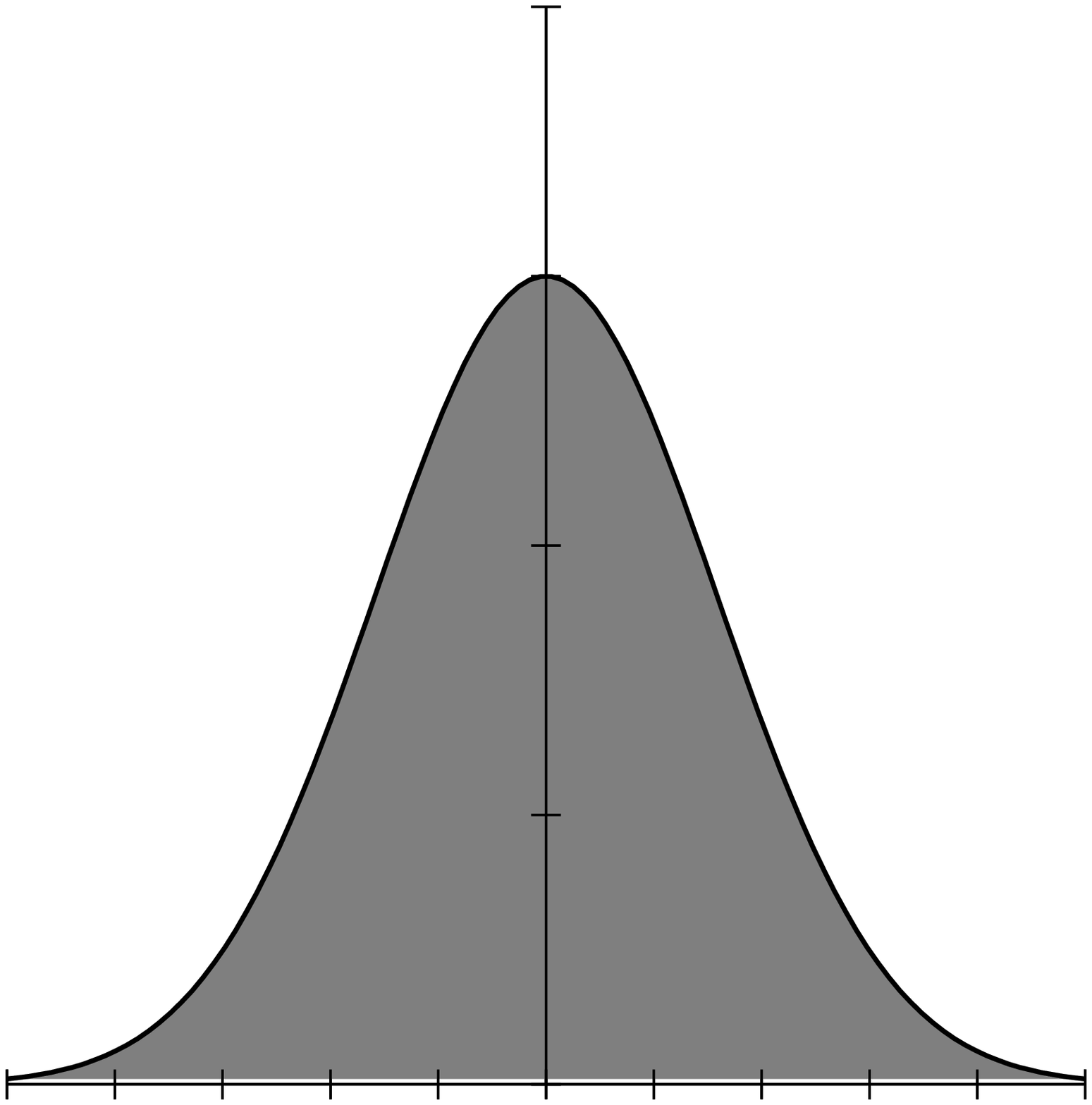}
	}
	}

	\put(-50,180){\(\times\)}

	\put(-10,240){
	\rotatebox{270}
	{
		\includegraphics[bb = 61 170 552 650,width=0.2\textwidth,clip=true]{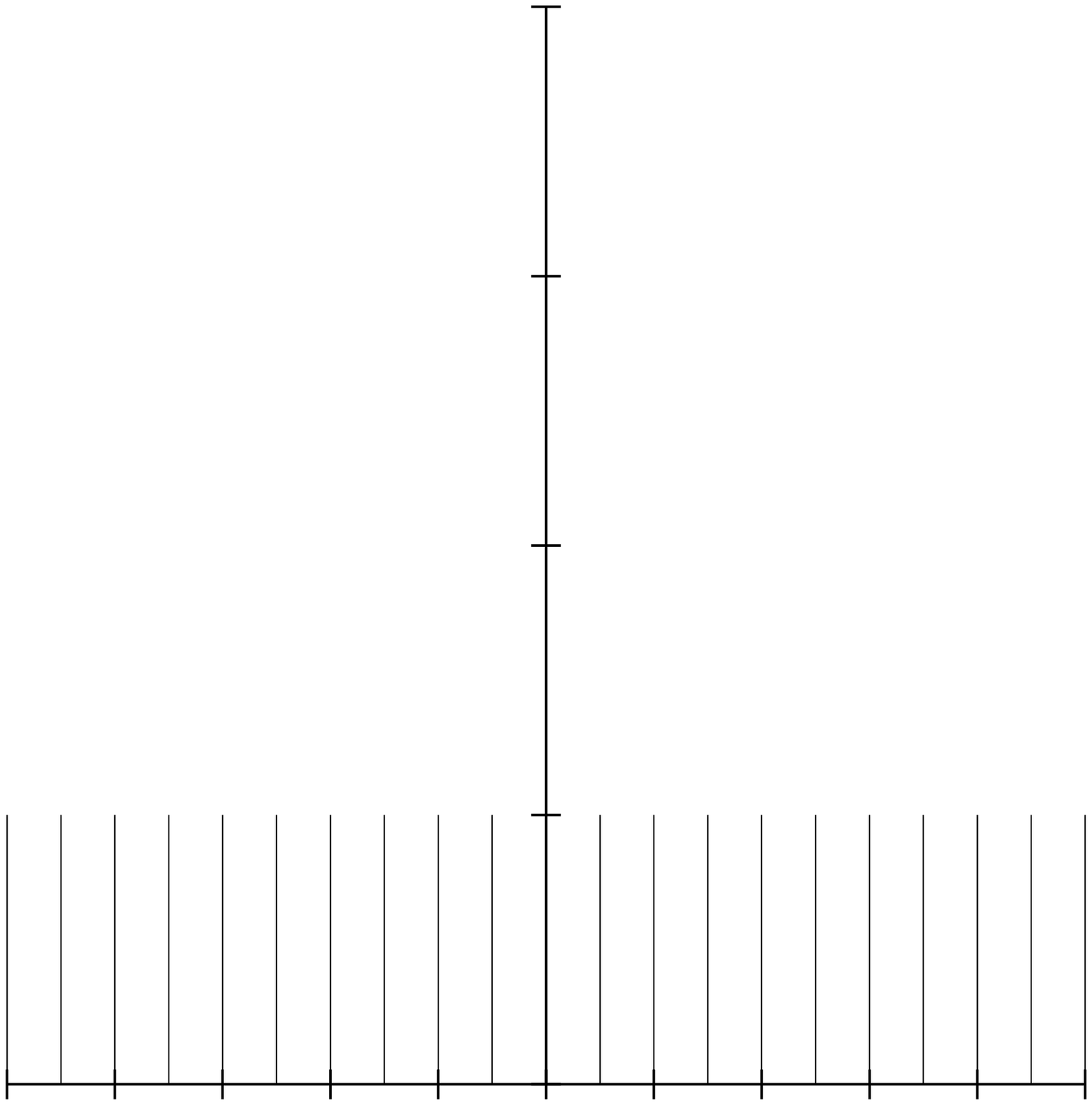}
	}
	}

	\put(130,180){\(=\)}

	\put(160,240){
	\rotatebox{270}
	{
		\includegraphics[bb = 61 170 552 650,width=0.2\textwidth,clip=true]{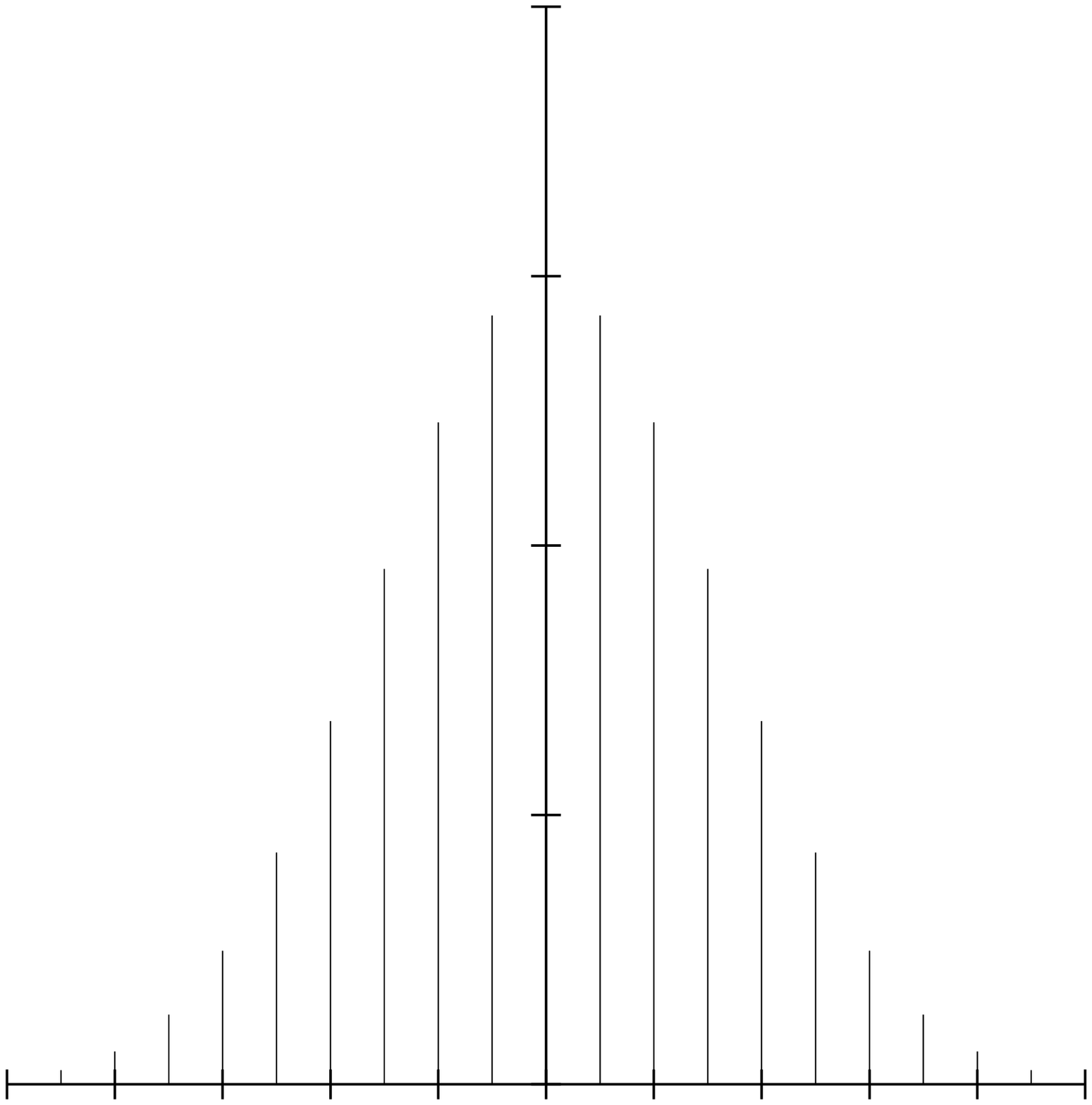}
	}
	}


	\put(-180,100){b)}

  	\put(-180,100){
	\rotatebox{270}
	{
		\includegraphics[bb = 61 170 552 650,width=0.2\textwidth,clip=true]{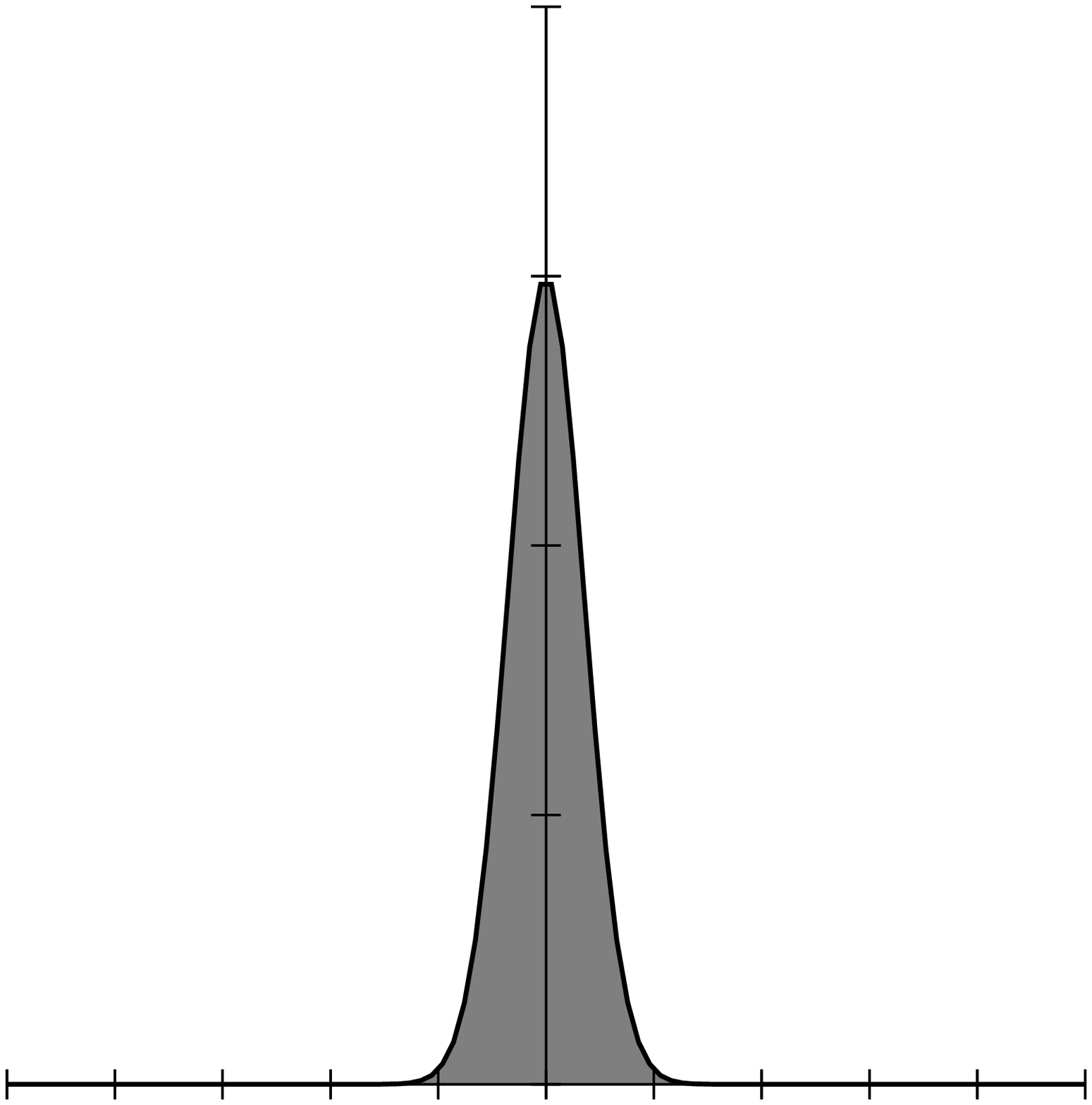}
	}
	}

	\put(-50,40){\(\circ\)}

	\put(-10,100){
	\rotatebox{270}
	{
		\includegraphics[bb = 61 170 552 650,width=0.2\textwidth,clip=true]{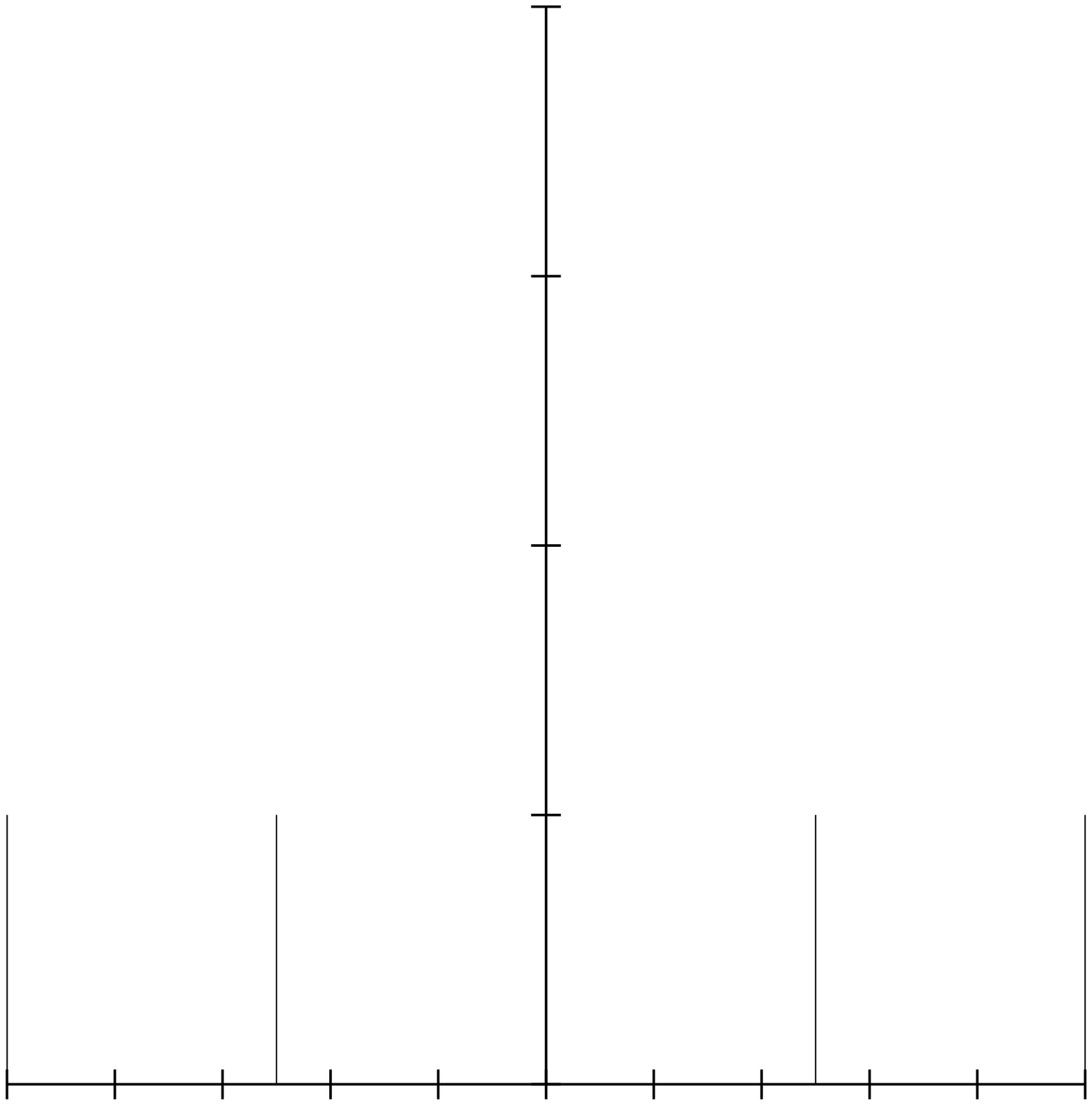}
	}
	}

	\put(130,40){\(=\)}

	\put(160,100){
	\rotatebox{270}
	{
		\includegraphics[bb = 61 170 552 650,width=0.2\textwidth,clip=true]{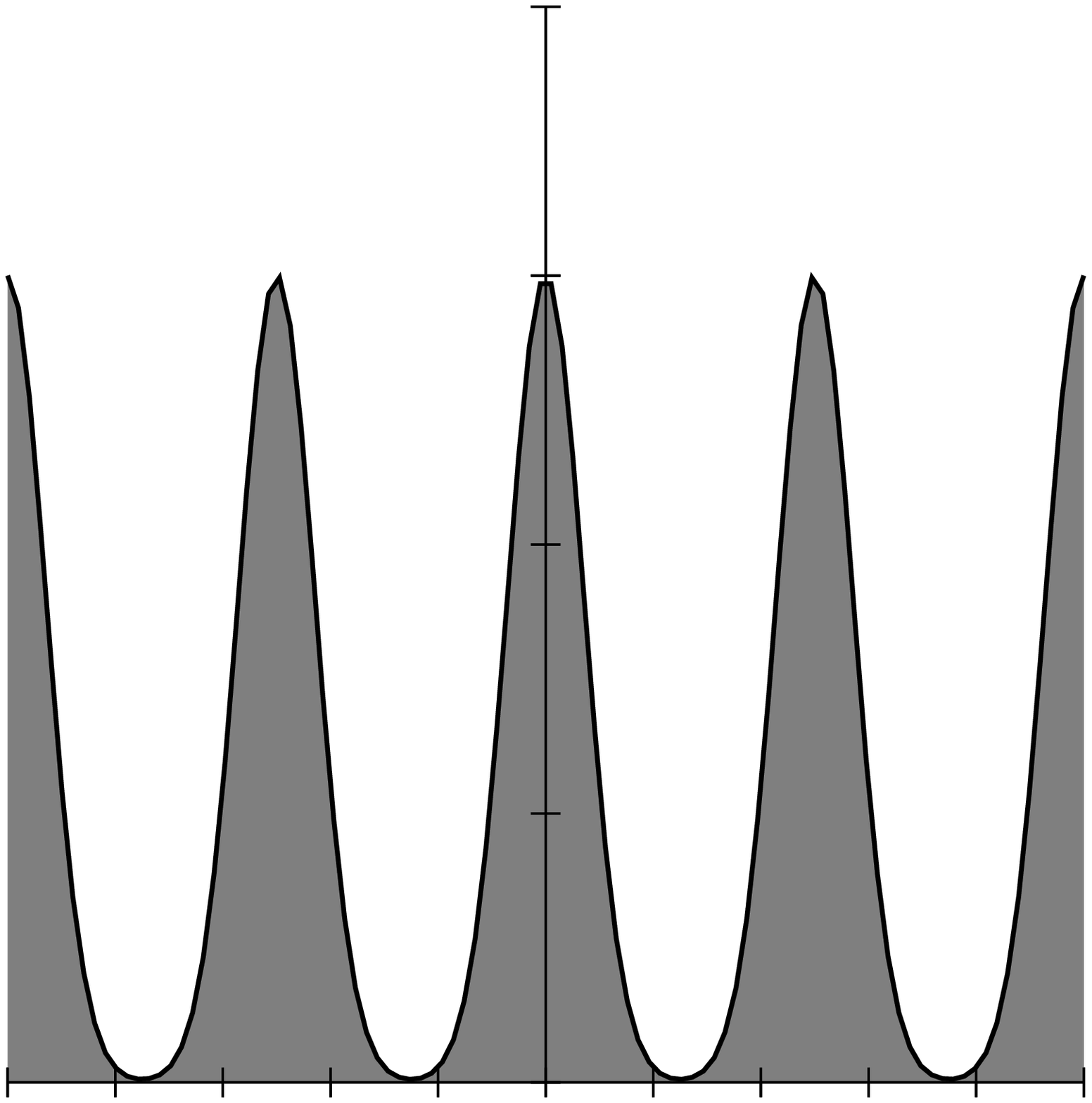}
	}
	}

        \end{picture}
	}
	\caption{ Multiplication a) and convolution b) of a function with the Dirac comb.}
	\label{fig:mult_conv_diraccomb}
\end{figure*}

\section{The requirements of FFTs}
\label{The_requirements_of_FFTs}
As already mentioned in the introduction, the FFT is a valuable tool in processing huge cosmological datasets due to its computational efficiency. This allows us to apply many mathematical operations like convolutions or deconvolutions to the data in an computational feasible way. However, using FFTs relies on two strong requirements as the function has to be discrete not only in real-space but also in Fourier-space.

It is a well known fact that a natural physical signal like the galaxy distribution is neither living in a discretized real nor a discretized Fourier-space. The reduction of such a signal to a set of finite and discrete sample points, thus, introduces an enormous loss of information.

As will be demonstrated below, the sampling theorem by Claude Shannon \citep{Shannon48,Shannon49} requires a function to be band limited, meaning, that the function has to be sufficiently smooth in order to have its support extending only up to a maximal frequency in Fourier-space. If this criterion would be fulfilled, then the complete information of the continuous signal can be represented by a set of discrete sample points in real-space. However, physical theories about the dark matter distribution and the power spectrum tell us, that the power spectrum possibly extends over all frequencies in Fourier-space, meaning that the matter and galaxy distributions cannot be sampled on a discrete grid without loss of information.

In addition, as already mentioned above, discreteness in real-space is not enough, and must be understood only as a necessary requirement for using FFTs.
Using FFTs additionally enforces the Fourier-space to be discrete. This is in agreement with the fact that the FFT requires the real-space signal to be periodic, meaning the signal can be represented by a finite set of Fourier waves.

For these reasons we claim the requirement of real-space and Fourier-space discreteness to be strong criteria in hindsight of information conservation, as in general no physical signal will fulfill these requirements in a natural way. 

In the following we will discuss how to optimally sample a continuous signal in order to conserve as much information as possible. In doing so, we will arrive at the ideal instrument response function of our computer, which allows us further intuitive insight into the problem of representing a real physical quantity on a grid of finite sample points.

\section{Discretizing the real-space}
\label{Descretize_Rspace}
As already mentioned a necessary requirement for the application of FFTs is the real-space discreteness of a signal. However, for the case discussed in this work the real physical signal is continuous in real-space and hence has to be discretized. This is usually achieved by dividing the continuous signal into small regions and associating a number with each of those. As this process also involves a loss of spatial resolution, and therefore requires to soften sharp features, information gets lost. This process of converting a continuous signal into a discrete representation can cause several artifacts like aliasing or Gibbs ringing.

A good way to understand, analyze and eliminate these artifacts is through Fourier analysis. In the following we will discuss the necessary requirements for real-space discretization, as was demonstrated by Claude Shannon in his sampling theorem \citep{Shannon48,Shannon49}.
\subsection{Sampling theorem}
\label{SAMPLING_THEOREM}
In order to discretize a continuous function, we will represent a point sample as a scaled Dirac impulse function.

With this definition, sampling a signal is equivalent to multiplying it by a grid of impulses, one at each sample point \(x_j=j\Delta x\), where \(j\) is an integer and \(\Delta x\) is the grid spacing \citep{Marschner1994}, as illustrated in Figure \ref{fig:mult_conv_diraccomb}.
 
Let \(f(x)\) be a continuous function, then its sampled version \(g(x)\) can be expressed by:
\begin{equation}
\label{eqn:Sampling_theorem1a}
g(x)=\Pi(x)\, f(x) \, ,
\end{equation}
where the sampling function \(\Pi(x)=\sum_{j=-\infty}^{\infty} \delta^D(x-j \Delta x)\) is a Dirac comb or impulse train. With this definition we yield the discrete function at the sample positions \(g_j\) as:
\begin{equation}
\label{eqn:discr_samples}
g_j \equiv f(j\Delta x) \, ,
\end{equation}

However, in order to demonstrate the origin of aliasing effects in the following we will focus on the Fourier transform of equation (\ref{eqn:Sampling_theorem1a}).
By making use of the convolution theorem \ref{convolution_theorem}, which states that the product of two functions in real-space yields a convolution in Fourier-space, the Fourier transform of the sampled function \(\hat{g}(p)\) can be expressed as: 
\begin{equation}
\label{eqn:Sampling_theorem2}
\hat{g}(p)=\left(\hat{\Pi} \circ \hat{f}\right)(p)  \, ,
\end{equation}
where the \(\hat{}\)-symbol denotes the Fourier transform of a function and the circle symbol \(\circ\) denotes a convolution.

As demonstrated in Appendix \ref{CFT_SAMPLING_OPERATOR} the Fourier transform of the Dirac comb \(\hat{\Pi}(p)\) is again a Dirac comb and is given by:
\begin{equation}
\label{eqn:CFT_Diraccomb}
\hat{\Pi}(p)= p_{s} \sum_{j=-\infty}^{\infty} \delta^D\left(p-j\,p_{s}\right) \, ,
\end{equation}
where \(p_{s}=2\, \pi / \Delta x\) is the repetition length in Fourier-space, which is often called the Nyquist rate.
Since \(\hat{\Pi}(p)\) is a Dirac comb the convolution in equation (\ref{eqn:Sampling_theorem2}) amounts to duplicating \(\hat{f}(p)\) at every point of \(\hat{\Pi}(p)\):
\begin{eqnarray}
\label{eqn:Aliasing_sum}
\hat{g}(p) &=& \left(\hat{\Pi} \circ \hat{f}\right)(p) \nonumber \\
&=& p_{s} \int_{-\infty}^{\infty} \sum_{j=-\infty}^{\infty} \delta^D\left(p'-j\,p_{s}\right) \hat{f}(p-p') dp' \nonumber \\&=& p_{s} \sum_{j=-\infty}^{\infty} \hat{f}\left(p-j\,p_{s}\right) \, ,
\end{eqnarray}
as displayed in Figure \ref{fig:mult_conv_diraccomb}.
This demonstrates that the Fourier representation of the sampled function \(\hat{g}(p)\) is a superposition of the true continuous Fourier transform \(\hat{f}(p)\) and all its replicas at positions \(j\, p_{s}\equiv j(2\pi)/\Delta x\). We call the copy of \(\hat{f}(p)\) at \(j=0\) the primary spectrum and all other copies alias spectra.
This result reveals that the sampling operation has left the original input spectrum \(\hat{f}(p)\) intact, just replicating it periodically in the Fourier domain with a spacing of \(p_{s}\) \citep{Wolberg1997}.

\begin{figure*}
	\centering
	{
	\begin{picture}(100,140)

\put(-140,140){a)}

  	\put(-140,140){
	\rotatebox{270}
	{
		\includegraphics[bb = 61 170 552 650,width=0.3\textwidth,clip=true]{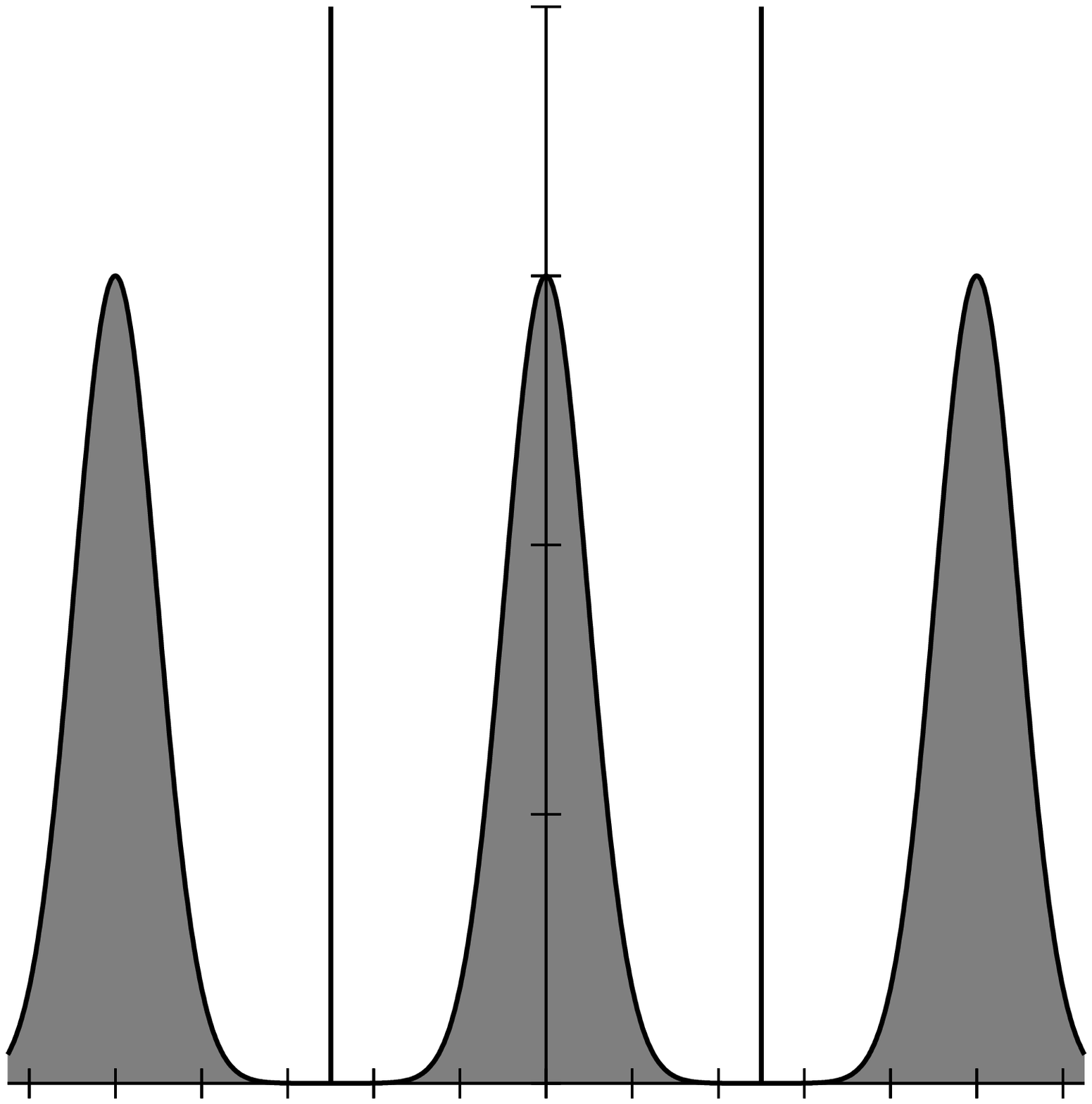}
	}
	}

\put(80,140){b)}
	
	\put(80,140){
	\rotatebox{270}
	{
		\includegraphics[bb = 61 170 552 650,width=0.3\textwidth,clip=true]{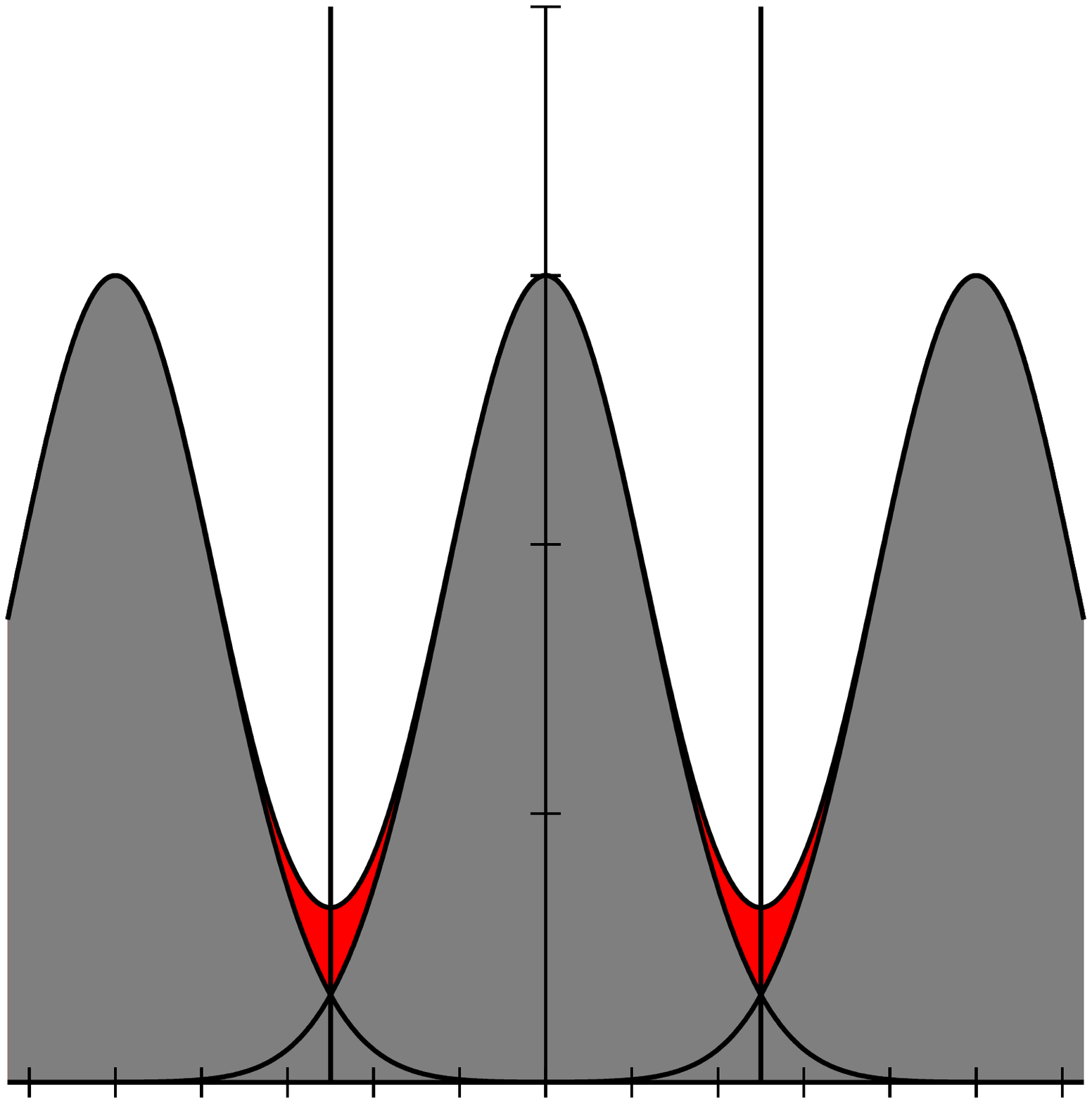}
	}
	}

        \end{picture}
	}
	\caption{a) Undistorted sampled function and b) sampled function with aliasing.}
	\label{fig:aliasing}
\end{figure*}

This suggests to rewrite the sampled spectrum \(\hat{g}(p)\) as a sum of two terms, the low-frequency (baseband), and the high frequency (aliasing) components;
\begin{eqnarray}
\label{eqn:Aliasing_sum}
\hat{g}(p) &=&  \hat{f}(p) + \sum_{j=-\infty, j \neq 0}^{\infty} \hat{f}\left(p-j p_{s}\right) \nonumber \\
&=&  \hat{f}(p) + \hat{F}_{HF}(p)
\end{eqnarray}
The baseband spectrum is exactly \(\hat{f}(p)\), and the high-frequency components, \(\hat{F}_{HF}(p)\), consists of the remaining replicated versions of \(\hat{f}(p)\) that constitute harmonic versions of the sampled signal \citep{Wolberg1997}.

A crucial observation in the study of sampled-data systems is that exact signal  reconstruction from the sampled data requires to discard the replicated spectra \(\hat{F}_{HF}(p)\), leaving only \(\hat{f}(p)\), the spectrum of the signal we seek to recover \citep{Wolberg1997}.

If the different replicas \(\hat{f}(p)\) do not overlap with the baseband spectrum, then we can recover \(\hat{f}(p)\) from \(\hat{g}(p)\), by simply multiplying \(\hat{g}(p)\)  with a function which is one inside the baseband and zero elsewhere, and therefore eliminates the high frequency contributions, see Figure \ref{fig:aliasing} a).
On the other hand, if the high-frequency contribution \(\hat{F}_{HF}(p)\) has some overlap with the baseband spectrum, there is no way to uniquely recover the original signal \(\hat{f}(p)\) from its sampled version \(\hat{g}(p)\), see Figure \ref{fig:aliasing} b).
Therefore, the only provision for exact sampling is that \(\hat{f}(p)\) must be undistorted due to the overlap with \(\hat{F}_{HF}(p)\) \citep{Wolberg1997}. For this to be true, two conditions must hold:
\begin{enumerate}
\item{The signal must be bandlimited. This avoids functions \(\hat{f}(p)\) with infinite extent that are impossible to replicate without overlap.}

\item{The sampling frequency \(p_{s} = (2\pi)/\Delta x\) must be greater than twice the maximum frequency \(p_{max}\) present in the signal. This can be understood by looking at Figure \ref{fig:aliasing}. This minimum frequency, known as the Nyquist rate, is the minimum distance between the spectra copies, each with bandwidth \(p_{s}\).}
\end{enumerate}

The first condition merely ensures that a sufficiently large sampling frequency \(p_{s}\) exists that can be used to separate replicated spectra from each other \citep{Wolberg1997}. The second condition provides an answer to the problem of the sufficiency of data samples to exactly recover the continuous input signal. It states that the signal can only be recovered exactly when \(p_{s} >2\,p_{max}\), with \(p_{max}=p_{Nyquist}\) being the Nyquist frequency, not to confuse with the Nyquist rate \citep{Wolberg1997}.

This is the sampling theorem as pioneered by Claude Shannon in his papers \citep{Shannon48,Shannon49}.

\subsection{Low-Pass-Filtering}
As seen in the previous section, when a signal is being sampled it must be band-limited if we are to recover its information content correctly. Natural signals, however, are not generally band-limited, and so must be low-pass filtered before they are sampled or equivalently, the sampling operation must include some form of local averaging. 

It is therefore required to apply a filter to the signal, which cuts away all Fourier modes higher than a certain maximal frequency \(p_{max}=p_{s}/2\). Such a filter is called an ideal low pass filter, and it's Fourier representation is given as:
\begin{eqnarray}
\label{eqn:Low_pass_filter}
\hat{W}(p) = \left \{ \begin{array}{ll}
  1 & \quad \mbox{for $|p|<p_{max}$}\\
  0 & \quad \mbox{for $|p|\geq p_{max}$}\\ \end{array} \right. \, .
\end{eqnarray}
It is ideal in the sense, that it has unity gain in the pass-band region, hence not introducing any attenuation of the Fourier modes to pass, and that it perfectly suppresses all the power in the stop-band region.
Applying this ideal low-pass filter to the signal by multiplication in Fourier-space yields the low-pass filtered signal:
\begin{equation}
\label{eqn:Filter_equation_Fspace}
\hat{h}(p)= \hat{W}(p)\,\hat{f}(p)  \, ,
\end{equation}
which according to the convolution theorem \ref{convolution_theorem} translates to its real-space representation as:
\begin{equation}
\label{eqn:Filter_equation_Rspace}
h(x)= \left(W\circ f\right)(x)  \, .
\end{equation}
The real-space representation \(W(x)\) of the ideal low pass filter is calculated in Appendix \ref{CFT_ideal_low_pass_filter} and yields a sinc function:
\begin{equation}
\label{eqn:sinc_ideal_Low_pass_filter_real_space}
W(x) = \frac{p_{max}}{\pi}\, sinc(p_{max}x) \, .
\end{equation}
Hence, low-pass filtering is equivalent to convolving with a sinc function in real-space.

As the sinc function has not such a sharp localization in space as a Dirac delta distribution, any sharp feature will be smeared out. In this sense low-pass filtering reduces spatial information, which cannot be uniquely restored.

The low-pass filtered function \(h(x)\) now meets the requirements of the sampling theorem, and can be discretized in real-space with a grid spacing \(\Delta x < \pi/p_{max}\). With these requirements the sampling procedure for a natural signal turns into:
\begin{equation}
\label{eqn:Sampling_Filter_equation_Rspace}
g(x)= \Pi(x)\, h(x) = \Pi(x)\, \left(W\circ f\right)(x)  \, .
\end{equation}
This process is graphically represented in Figure \ref{fig:sampling_scheme} a).

\begin{figure*}
	\centering
		\rotatebox{270}
	{
		\includegraphics[bb = 61 153 541 650,width=0.5\textwidth,clip=true]{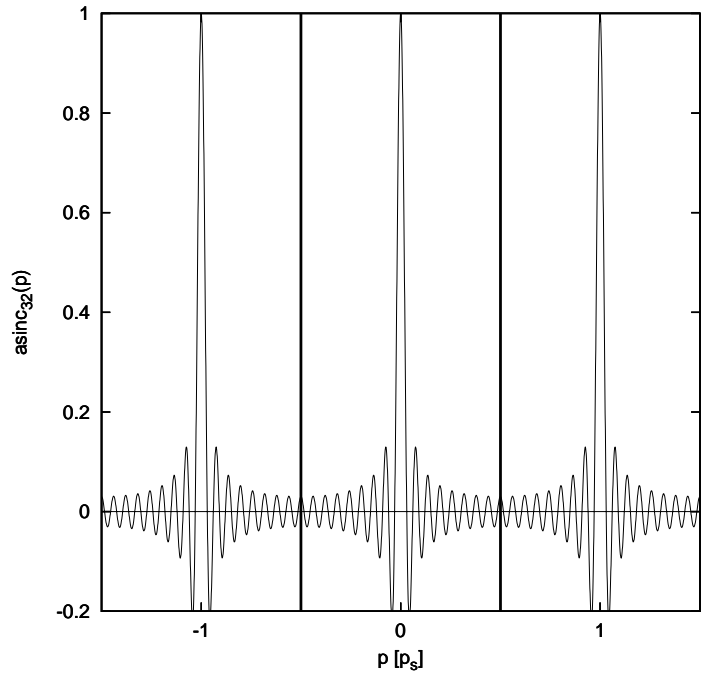}
	}
	\caption{Aliased sinc function for N=32. The vertical lines limit the base-band.}
	\label{fig:aliased_sinc}
\end{figure*}

\subsection{Sampling on a finite real-space domain}
\label{Real_space_sampling_with_finite_sums}
We have demonstrated above, that a band-limited signal can be uniquely recovered from its discretized approximation. This is due to the fact that there exists a sufficiently high sampling frequency to separate the aliased copies from the base-band spectrum and that the sampling operation has left the original spectrum \(\hat{f}(p)\) intact.
However, this is closely related to the fact that the samples are distributed over the infinite domain of real-space as the sum in the sampling operator \(\Pi(x)\) extends from minus to plus infinity since only then the Fourier transform of the sampling operator is again a Dirac comb.
Unfortunately, this is not the case in real world applications, especially since it is generally not possible to evaluate the infinite sampling sum with a computer. Therefore, one generally wants to restrict the investigation to a finite subset of samples. For technical reasons one usually chooses a rectangular sub domain. Therefore the sampling operator can be split in a zero centered sub domain and the surrounding rest as follows:
\begin{eqnarray}
\label{eqn:split_sampling_operator}
\Pi(x) &=&\sum_{j=-\infty}^{\infty} \delta^D(x-j \Delta x) \nonumber \\ 
&=& \sum_{j=-\frac{N}{2}}^{\frac{N}{2}} \delta^D(x-j \Delta x) +\sum_{j=\frac{N}{2}+1}^{\infty} \left[\delta^D(x-j \Delta x) + \delta^D(x+j \Delta x) \right ] \nonumber \\
&=& \Pi_{N}(x) + \Xi_{N}(x) \, ,
\end{eqnarray}
where we introduced the finite sampling operator \(\Pi_{N}(x)\) and the trans-domain sampling operator (Jordan operator) \(\Xi_{N}(x)\). The sampled signal on the sub-domain is then given by:
\begin{equation}
\label{eqn:sub_domain_samples}
g_{N}(x)= g(x) - R_{N}(x)\, ,
\end{equation}
with \(g_{N}(x)=\Pi_{N}(x) f(x)\) and the rest term \(R_{N}(x)=\Xi_{N}(x) f(x)\). As demonstrated above \(g(x)\) contains all information to uniquely recover the band-limited signal. Thus, by reducing the set of signal samples to a finite subset information gets lost and the signal cannot be reconstructed uniquely. This can be easily seen by Fourier analysis.

According to Appendix \ref{Finite_sum_sampling_operator} the Fourier transform of the finite sampling operator is given as:
\begin{equation}
\label{eqn:FT_finite_sampling_op}
\hat{\Pi}_{N}(p)= \rm{asinc}_{N}(p) \, ,
\end{equation} 
with the aliased sinc function \(\rm{asinc}_{N}(p)\) being: 
\begin{equation}
\label{eqn:aliased_sinc_function}
\rm{asinc}_{N}(p) =  \frac{\sin \left (\frac{p\Delta x}{2}(N+1)\right)}{\sin\left(\frac{p\Delta x}{2}\right)} \, .
\end{equation} 
The aliased sinc function is plotted in figure \ref{fig:aliased_sinc}. It is obvious that the Fourier transform of the finite sampling operator is not a Dirac comb anymore, though it still exhibits the property of having resonance peaks at the repetition length \(p_s\). Sampling on a finite domain therefore amounts to convolving the Fourier spectrum of the true continuous signal with the aliased sinc function. This operation however does not leave the base-band spectrum intact, and, unlike as in the Sampling theorem, unique recovery of the underlying signal is not possible anymore.
It is also obvious that due to the shape of the aliased sinc function the base-band spectrum will always be affected by the higher order spectra, which are overlapping with the baseband.
However, low pass filtering will not completely eliminate these aliasing contributions but it can alleviate them.

Therefore, in general it is not possible to uniquely restore information from signals which have been sampled on finite domains, and interpretations drawn from these sampled signals will always be afflicted with some uncertainty. 

\section{Discretizing The Fourier-space}
\label{DISCRETIZING_FOURIERSPACE}
In the previous section we discussed the theory of discretizing the real-space representation of a continuous signal. We demonstrated that in real world applications due to the finite sampling operator the sampling theorem is not applicable anymore in a strict sense. However, discretizing the real-space representation of the signal is usually not enough for data processing purposes. It can for instance be easily demonstrated that the Fourier transform of such a discrete signal is still a continuous function in Fourier-space. Nevertheless, especially applications in which FFT techniques are used require also discrete Fourier-space representations of the signal. Therefore, in the following we will derive a sampling method, designed to be used with FFTs.

\subsection{FFTs and the Fourier-space representation}
\label{FFTs_and_the_Fourier_space_representation}

FFTs, since they are based on the Fourier-series, require the function to be periodic on the observational interval \(L\). This assumes that the observed signal can be decomposed into eigen-modes of a resonator with length \(L\). This resonator has a ground frequency of \(p_0=2\pi/L\) and all other harmonic frequencies can be obtained by multiplying with an integer \(p_i=p_0\, i\). Thus, the Fourier domain observation obtained via an FFT is discrete. It is worth mentioning, that this discreteness in Fourier-space arises from applying the FFTs and therefore implicitly assuming the function to be periodic on the observational interval. In doing so we introduced a second discretizing process, but this time the Fourier-space is being discretized. Thus, FFTs  should be considered as an additional filter for the true underlying signal.

As a consequence the question arises how the Fourier modes obtained by an FFT of the observed signal are related to the Fourier modes of the true natural signal. This is of special interest in the case of cosmological matter power spectrum estimation. To demonstrate this we will follow the usual ideal approach to discretize a signal, by first low-pass filtering and then sampling it to grid positions. The filtered and gridded signal can then be represented by:
\begin{equation}
\label{eqn:filtered_and_gridded}
g_j =\int_{-\infty}^{\infty} dx \,W(\Delta x \, j-x)\, f(x) \, ,
\end{equation}
where \(W(x)\) is the ideal low-pass filter as given by equation (\ref{eqn:sinc_ideal_Low_pass_filter_real_space}).
Applying the FFT as defined in Appendix \ref{Discrete_Fourier_transformation} yields:
\begin{eqnarray}
\label{eqn:FFT_filtered_and_sampled}
\!\! \hat{g}_k\!\!&=&\!\!\! \hat{C}\sum_{j=0}^{N-1}g_j\,e^{-2\pi j k \frac{\sqrt{-1}}{N}} \nonumber \\
\!\!\!\!&=&\!\!\! \hat{C}\sum_{j=0}^{N-1}\,e^{-2\pi j k \frac{\sqrt{-1}}{N}}\,\int_{-\infty}^{\infty} dx \,W(\Delta x \, j - x)\, f(x) \nonumber \\
\!\!\!\!&=&\!\!\! \frac{\hat{C}}{2\pi}\int_{-\infty}^{\infty} dp \hat{f}(p)\,\sum_{j=0}^{N-1}\,e^{-2\pi j k \frac{\sqrt{-1}}{N}}\,\int_{-\infty}^{\infty} dx \,W(\Delta x \, j - x)\, e^{\sqrt{-1}\,p\,x} \nonumber \\
\!\!\!\!&=&\!\!\! \frac{\hat{C}}{2\pi}\int_{-\infty}^{\infty} dp\, \hat{f}(p)\,\sum_{j=0}^{N-1}\,e^{-2\pi j k \frac{\sqrt{-1}}{N}}\, \left(e^{\sqrt{-1}\,p\,\Delta x j}\right)\,\int_{-\infty}^{\infty}\!\!\!\!\! dx' \,W(x')\, e^{-\sqrt{-1}\,p\,x'} \nonumber \\
\!\!\!\!&=&\!\!\! \frac{\hat{C}}{2\pi}\int_{-\infty}^{\infty} dp\, \hat{f}(p) \hat{W}(p)\,\sum_{j=0}^{N-1}\,e^{-2\pi j k \frac{\sqrt{-1}}{N}}\, \left(e^{\sqrt{-1}\,p\,\Delta x j}\right) \nonumber \\
\!\!\!\!&=&\!\!\! \frac{1}{2\pi}\int_{-\infty}^{\infty} dp\, \hat{f}(p)\, \hat{W}(p)\,\hat{C}\,\sum_{j=0}^{N-1}\,e^{-2\pi j k \frac{\sqrt{-1}}{N}}\, \left(e^{\sqrt{-1}\,p\,\Delta x j}\right) \nonumber \\
\!\!\!\!&=&\!\!\! \frac{1}{2\pi} \int_{-\infty}^{\infty} dp\, \hat{f}(p)\, \hat{W}(p)\,\hat{C}\,\sum_{j=0}^{N-1}\,\left(e^{-\sqrt{-1} \Delta x j (\Delta p k -p)}\right) \nonumber \\
\!\!\!\!&=&\!\!\! \frac{1}{2\pi} \int_{-\infty}^{\infty} dp\, \hat{f}(p)\, \hat{W}(p)\, \hat{U}(\Delta p k - p)  \, , \nonumber \\
\end{eqnarray}
where we have introduced the Fourier-space lattice interval \(\Delta p=2\pi/(\Delta x N)=2\pi/L\) and the mode coupling function \(\hat{U}(\Delta p k - p)\):
\begin{eqnarray}
\label{eqn:mode_coupling_function}
\hat{U}(\Delta p k - p) &=&\hat{C}\,\sum_{j=0}^{N-1}\,e^{-\sqrt{-1} \Delta x j (\Delta p k -p)} \nonumber \\
&=&\hat{C}\,\sum_{j=0}^{N-1}\,\left(e^{-\sqrt{-1} \Delta x (\Delta p k -p)}\right )^j \nonumber \\
&=&\hat{C}\,\frac{1-e^{-\sqrt{-1} \Delta x (\Delta p k -p)\,N}}{1-e^{-\sqrt{-1} \Delta x (\Delta p k -p)}} \nonumber \\
&=&\hat{C}\,e^{-\sqrt{-1} \frac{\Delta x}{2} (\Delta p k -p)\,(N-1)}\,\frac{e^{\sqrt{-1} \frac{\Delta x}{2} (\Delta p k -p)\,N}-e^{-\sqrt{-1} \frac{\Delta x}{2} (\Delta p k -p)\,N}}{e^{\sqrt{-1} \frac{\Delta x}{2} (\Delta p k -p)}-e^{-\sqrt{-1} \frac{\Delta x}{2} (\Delta p k -p)}} \nonumber \\
&=&\hat{C}\,e^{-\sqrt{-1} \frac{\Delta x}{2} (\Delta p k -p)\,(N-1)}\,\frac{\sin \left(\frac{\Delta x}{2} (\Delta p k -p)\,N \right)}{\sin \left(\frac{\Delta x}{2} (\Delta p k -p) \right)} \nonumber \\
&=&\hat{C}\,e^{-\sqrt{-1} \frac{\Delta x}{2} (\Delta p k -p)\,(N-1)}\, asinc(\Delta p k -p) \, .
\end{eqnarray}
This immediately demonstrates that the spectral representation obtained by the FFT is a filtered version of the true Fourier transform of the signal.
It can be easily seen, that if the true signal is periodic on the observational domain \(L\) then the convolution with the filter kernel \(\hat{U}(p)\) vanishes. Hence, in this case we obtain the true Fourier representation of the signal.
However, as natural signals are in general not periodic on the observational domain they cannot be decomposed in a discrete set of Fourier waves, and hence some Fourier smoothing must be introduce, in order to represent the non-periodic function by a periodic function.
It is also worth noticing, that the mode coupling function does not only affect the amplitude of certain modes, but also introduces a phase shifting factor. This is due to the fact, that the FFT applies a window which is not centered on the origin to limit the real-space domain.
For these reasons it is not possible to uniquely recover the true Fourier spectrum from the FFT Fourier representation of the signal.

\begin{figure*}
	\centering
	{
	\begin{picture}(100,140)

\put(-210,140){a)}

  	\put(-210,140){
	\rotatebox{270}
	{
		\includegraphics[bb = 61 170 552 670,width=0.28\textwidth,clip=true]{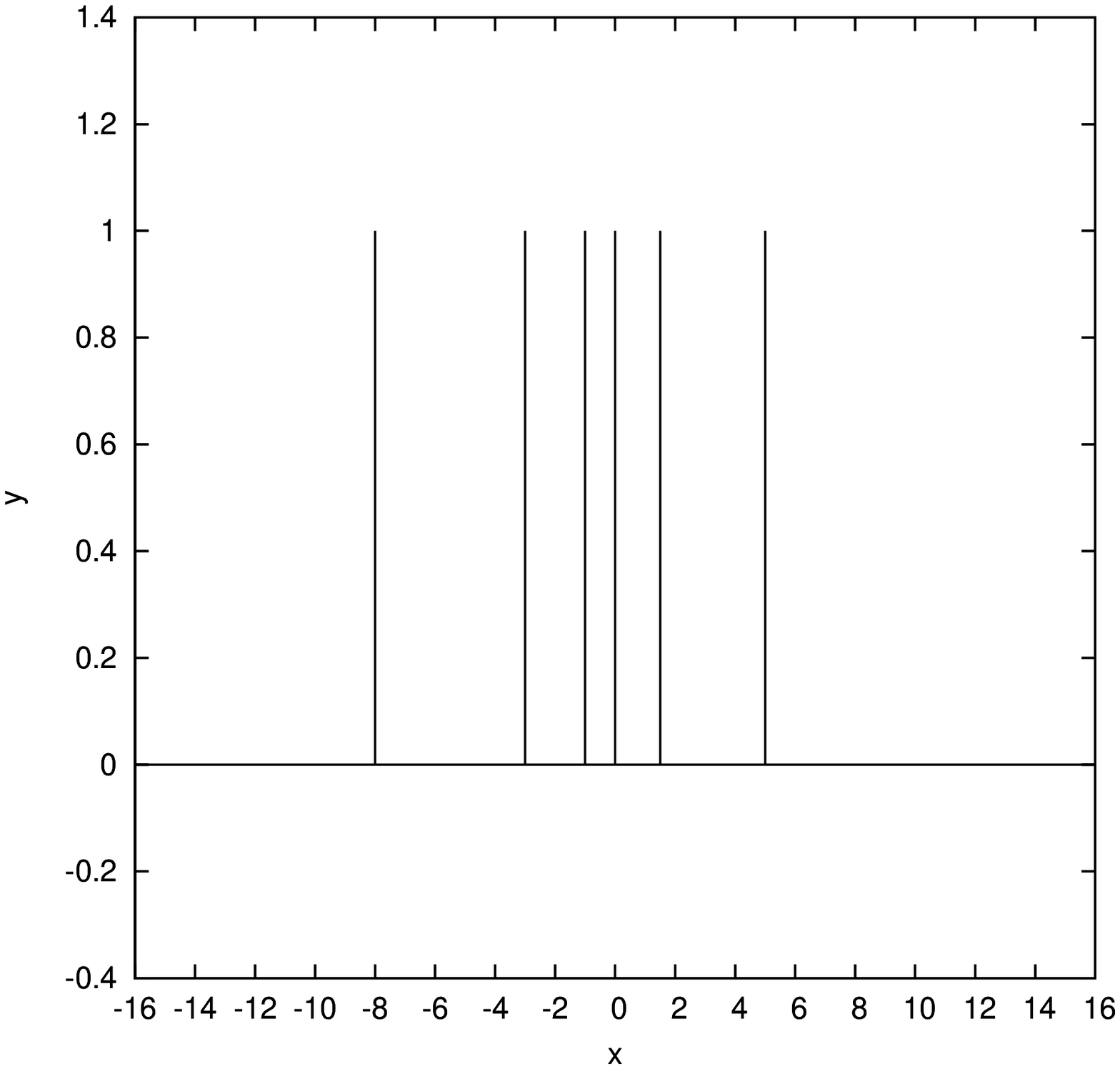}
	}
	}

\put(-20,140){b)}	
	\put(-20,140){
	\rotatebox{270}
	{
		\includegraphics[bb = 61 170 552 670,width=0.28\textwidth,clip=true]{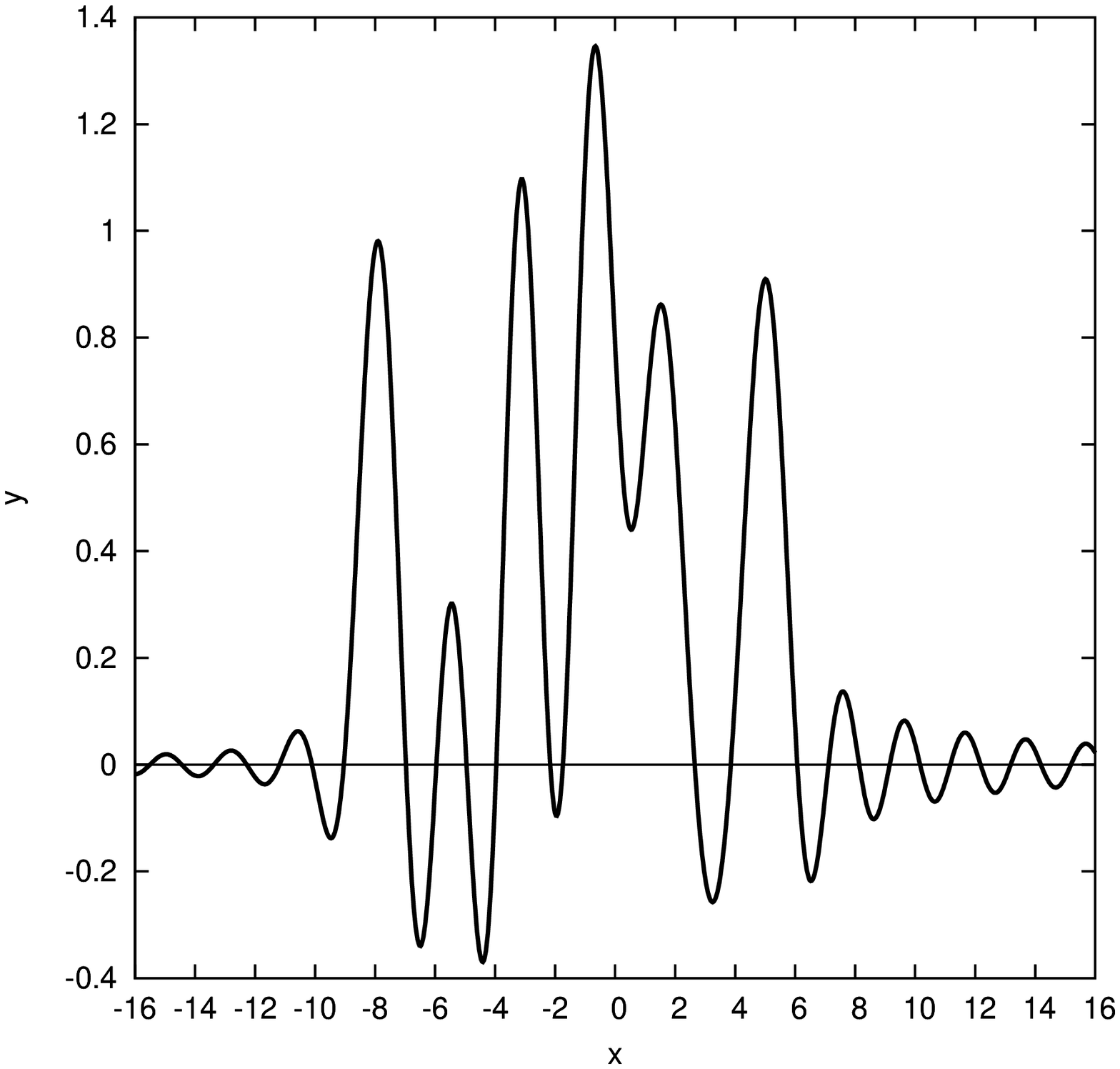}
	}
	}

\put(170,140){c)}	
	\put(170,140){
	\rotatebox{270}
	{
		\includegraphics[bb = 61 170 552 670,width=0.28\textwidth,clip=true]{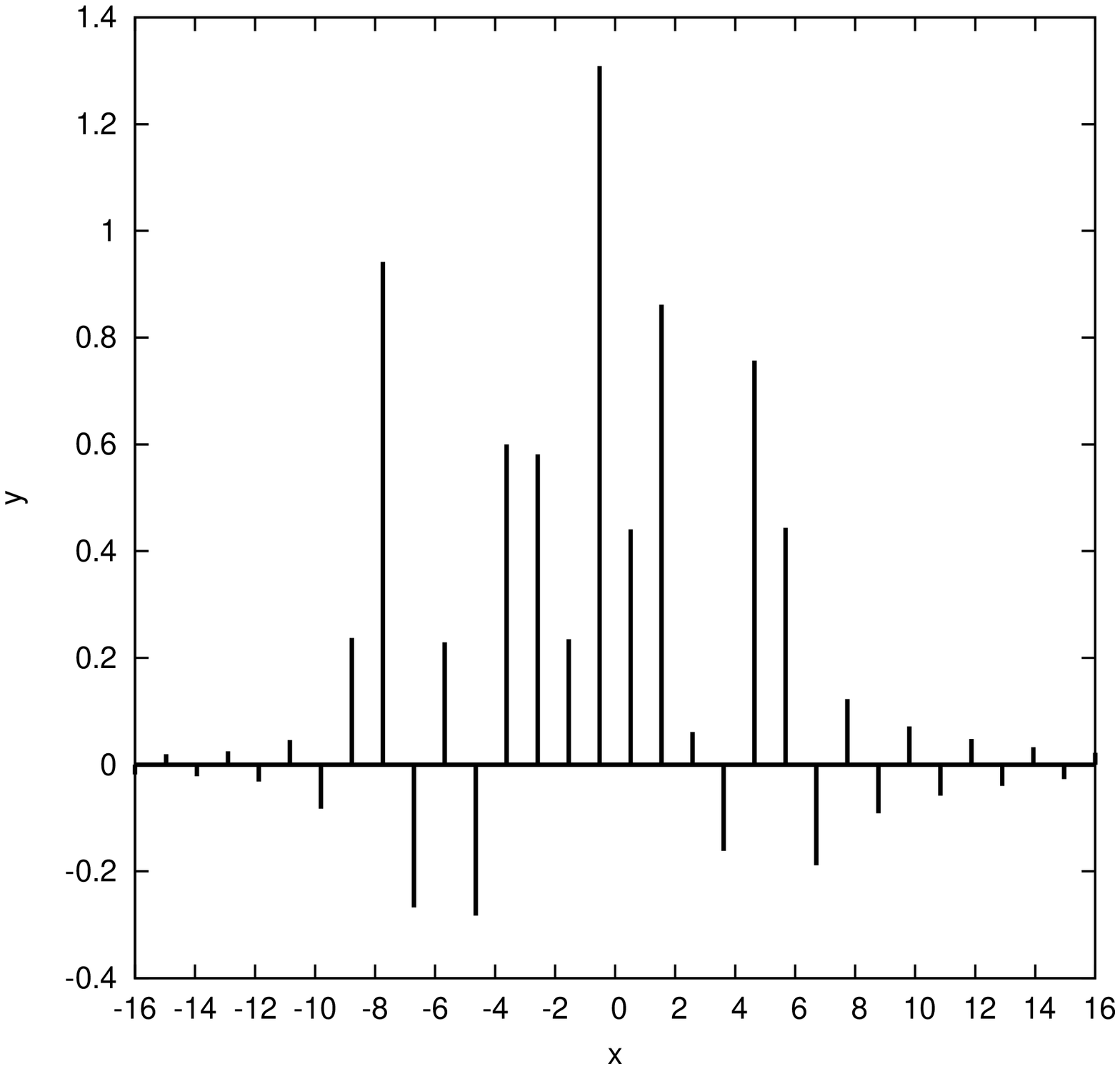}
	}
	}

        \end{picture}
	}
	\caption{Sampling process of a galaxy distribution. The galaxy distribution displayed in panel a) is low-pass filtered in panel b) and finally sampled at discrete positions in panel c).}
	\label{fig:model_sampling}
\end{figure*}
\subsection{Sampling in Fourier-space}
In the previous section we demonstrated that applying an FFT or in more general a DFT to a low pass filtered non-periodic signal yields a Fourier representation, which deviates from the true Fourier representation by a convolution with the mode coupling function \(\hat{U}(p)\).
The continuous signal therefore has to be filtered in such a way that its Fourier representation has a discrete spectrum. This can be achieved by sending the signal through a resonator of length \(L\), which effectively means the convolution with a Dirac comb in real-space. The proof of this can trivially been shown by making use of the convolution theorem, and knowing that the Fourier transform of the Dirac comb is again a Dirac comb.
We introduce the realspace replication operator \(\Pi_R(x)\) as
\begin{equation}
\label{eqn:Replication_Operator}
\Pi_R(x) =\sum_{j=-\infty}^{\infty} \delta^D(x-j L)\, ,  
\end{equation}
with its Fourier transform, according to Appendix (\ref{CFT_SAMPLING_OPERATOR}), being:
\begin{equation}
\label{eqn:Replication_Operator_FS}
\hat{\Pi}_R(p) =\frac{2\,\pi}{L}\sum_{j=-\infty}^{\infty} \delta^D\left(p-j\, \Delta p\right)\, .  
\end{equation}
The replication operator allows us therefore to formulate the necessary step of discretizing the Fourier-space representation of the continuous signal, by evaluating the following convolution:
\begin{equation}
\label{eqn:Replicated_function}
f_R(x) =\left(\Pi_R\circ f\right)(x)\, ,  
\end{equation}
where \(f_R(x)\) is the replicated signal in real-space. The according Fourier space representation then reads:
\begin{equation}
\label{eqn:Replicated_function_FS}
\hat{f}_R(p) =\hat{\Pi}_R(p)\, \hat{f}(p) = \frac{2\,\pi}{L}\sum_{j=-\infty}^{\infty} \delta^D\left(p-j\, \Delta p\right) \hat{f}(p)\, .  
\end{equation}
This immediately demonstrates that \(\hat{f}_R(p)\) is discrete in Fourier-space.
Forcing a continuous non-periodic function to be periodic, or sending it through a resonator, is a filtering process, which discretizes the Fourier space representation.
We can now replace \(\hat{f}(p)\) by the discretized function \(\hat{f}_R(p)\) in equation (\ref{eqn:FFT_filtered_and_sampled}) to yield:
\begin{eqnarray}
\label{eqn:FFT_filtered_and_sampled_discrete_Fspace}
\hat{g}_k&=& \frac{1}{2\pi} \int_{-\infty}^{\infty} dp\, \hat{f}_R(p)\, \hat{W}(p)\, \hat{U}(\Delta p k - p)\nonumber \\ 
&=& \frac{1}{2\pi} \int_{-\infty}^{\infty} dp\, \frac{2\,\pi}{L}\sum_{l=-\infty}^{\infty} \delta^D\left(p-l\, \Delta p\right) \hat{f}(p)\, \hat{W}(p)\, \hat{U}(\Delta p k - p)\nonumber \\
&=& \frac{1}{L}\sum_{l=-\infty}^{\infty} \hat{f}(l\, \Delta p)\, \hat{W}(l\, \Delta p)\, \hat{U}(\Delta p\, (k - l))\, .
\end{eqnarray}
Note, that the mode coupling function now only depends on two integers, and therefore is a matrix in discrete Fourier-space. The mode coupling function is then:
\begin{eqnarray}
\label{eqn:mode_coupling_function_disc}
\hat{U}(\Delta p\, (k - l)) &=&\hat{C}\,\sum_{j=0}^{N-1}\,e^{-\sqrt{-1} \Delta x \Delta p j (k -l)} \nonumber \\
&=&\hat{C}\,\sum_{j=0}^{N-1}\,e^{- 2 \pi\,\sqrt{-1} j \frac{(k -l)}{N}} \nonumber \\
&=&\hat{C}\,N\,\delta^K_{lk} \, ,
\end{eqnarray}
where for the last equality we refer to Appendix \ref{DISC_MODE_COUPLING_FUNC}.

Using this result in equation (\ref{eqn:FFT_filtered_and_sampled_discrete_Fspace}) then yields:
\begin{equation}
\label{eqn:FFT_filtered_and_sampled_discrete_Fspace_a}
\hat{g}_k=\frac{\hat{C} \, N}{L} \hat{f}(k\, \Delta p)\, \hat{W}(k\, \Delta p)\, .
\end{equation}
This is a remarkable result. As the ideal low-pass filter \(\hat{W}(p)\), as given by equation (\ref{eqn:Low_pass_filter}), has unity gain in the pass band, it is possible to uniquely recover individual Fourier modes of the true physical continuous and non-periodic signal.

It is therefore clear, that the use of FFT techniques requires to sample the function twice, once in real-space, and once in Fourier-space. This can simply be achieved by replicating the continuous and non-periodic signal and low pass filtering it. According to Appendix \ref{Ideal_discretization_kernel} the ideal discretization filter kernel \(\Psi(x)\) would then be an aliased sinc function:
\begin{equation}
\label{eqn:ideal_discretization_kernel}
\Psi(x)=\frac{1}{L} \rm{asinc}_{N}(x)\, .
\end{equation}
This filtering procedure can be accomplished in a two step filtering process, where one first applies the replication operator and then the low-pass filter, or vice versa, to the continuous non-periodic signal. From this two step filtering it is also clear that there exist two parameters to adjust the sampled field. The first parameter is the cut-off frequency \(p_{max}\) of the low-pass filter, which controls the real-space resolution, the second parameter is the resonator length \(L\) which controls the Fourier-space resolution, or the spatial domain under consideration.

\subsection{The instrument response function of our computer and the loss of information}
\label{Instrument_Response}
In the previous sections we demonstrated that in order to process information via FFT techniques on a computer the function has to be band limited and discrete in Fourier-space meaning the function has to be periodic on the observational interval and can be represented by only a finite amount of Fourier waves.
As natural signals in general do not obey these requirements processing the data with a computer has to be understood as an additional observational and filtering step, which modifies the input data.

Therefore, in order to make inferences from the information stored on a computer towards the information about a real physical observable, one must take into account the systematics introduced by the sampling process and the use of FFTs or DFTs.
The discrete representation on a computer is only an approximation to the real continuous signal, and hence cannot represent all properties of the true original physical observable. This will be demonstrated now by the case of the strictly positive galaxy density field.

\begin{figure*}
	\centering
	{
	\begin{picture}(100,140)

\put(-140,140){a)}

\put(20,80)
{
\begin{picture}(100,140)
\put(-160,38){\(f(x)\)}
\put(-140,40){\vector(1,0){40}}
\put(-92,38){\(\circ\)}
\put(-90,40){\circle{12}}
\put(-90,10){\vector(0,1){20}}
\put(-103,0){filtering}
\put(-80,40){\vector(1,0){40}}
\put(-33,37){\(\Pi\)}
\put(-30,40){\circle{12}}
\put(-30,10){\vector(0,1){20}}
\put(-43,0){sampling}
\put(-20,40){\vector(1,0){40}}
\put(26,38){\(g_j\)}
\end{picture}
}

\put(-140,60){b)}
\put(20,0)
{
\begin{picture}(100,140)
\put(-160,38){\(f(x)\)}
\put(-140,40){\vector(1,0){40}}
\put(-92,38){\(\circ\)}
\put(-90,40){\circle{12}}
\put(-90,10){\vector(0,1){20}}
\put(-103,0){prefiltering}
\put(-80,40){\vector(1,0){40}}
\put(-33,37){\(\Pi\)}
\put(-30,40){\circle{12}}
\put(-30,10){\vector(0,1){20}}
\put(-43,0){supersampling}
\put(-20,40){\vector(1,0){40}}
\put( 28,38){\(\circ\)}
\put(30,40){\circle{12}}
\put(30,10){\vector(0,1){20}}
\put(17,0){filtering}
\put(40,40){\vector(1,0){40}}
\put(87,37){\(\Pi\)}
\put(90,40){\circle{12}}
\put(90,10){\vector(0,1){20}}
\put(70,0){downsampling}
\put(100,40){\vector(1,0){40}}
\put(146,38){\(g_j\)}
\end{picture}
}
        \end{picture}
	}
	\caption{Ideal sampling scheme a) and two stage supersampling scheme b).}
	\label{fig:sampling_scheme}
\end{figure*}
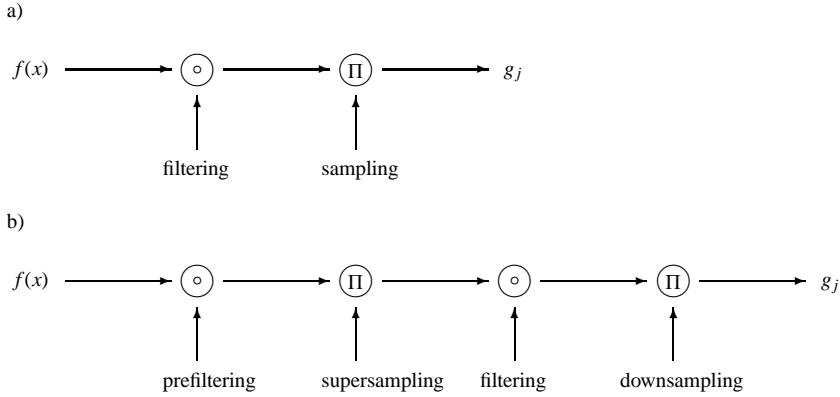
In many cosmological applications, like for instance power spectrum estimation, galaxies are considered to be point particles represented by a Dirac delta distribution. With this definition a galaxy distribution consisting of \(M\) galaxies can be expressed as:
\begin{equation}
\label{eqn:Gal_dista}
\rho(x)= \sum_{p=1}^{M} \delta^D(x-x_p) \, ,
\end{equation}
with \(x_p\) being the position of a galaxy. As can be seen easily this density field is strictly positive by definition.
However, applying the ideal low pass filter yields:
\begin{eqnarray}
\label{eqn:low_pass_filtered_Gal_dist}
\rho'(x)&=&\int_{-\infty}^{\infty} dx'\,W(x-x')\, \rho(x') \nonumber \\
&=& \sum_{p=1}^{M} \,W(x-x_p)\, .
\end{eqnarray}
This can be understood as replacing the Dirac delta distribution by the ideal low-pass filter kernel and therefore giving each galaxy the shape of a sinc function.

In Figure \ref{fig:model_sampling} we display a simple example of a low-pass filtered galaxy distribution with positions \(x_p \in \{-8,-3,-1,0,1,5\}\) to be sampled on a grid with grid spacing \(\Delta x = 1\). As already suggested above, the low-pass filtering introduces some sort of spatial smoothing due to the finite width of the sinc kernel. Hence, two galaxies being closer together than the grid spacing cannot be resolved independently as can be seen in the cases of the galaxies at position \(x_p=-1\) and \(x_p=0\). Another important thing to remark is that due to the oscillatory nature of the sinc function, the superposition of sinc functions will lead to positive and negative interference. This might introduce density peaks at positions where no peak would be observed in the true natural signal, which is demonstrated in Figure \ref{fig:model_sampling} at position \(x=-5.5\). 

As the sinc function is not strictly positive, the low pass filtered galaxy density given in equation (\ref{eqn:low_pass_filtered_Gal_dist}) does not possess the physical property of being strictly positive as the original observable, demonstrated as well by Figure \ref{fig:model_sampling}.
This property will only be restored in the limit of infinite resolution, when the sinc function approaches the Dirac delta distribution.

Being a strictly positive density field thereby is a physical property which cannot be represented by the sampled galaxy density field.
This result will also be true for other physical properties which will only be recovered in the limit towards infinite resolution.

However, it is worthwhile mentioning that though the galaxy density field may have negative contributions all integral quantities like total number of galaxies or total mass, are identical to the ones which could be obtained by integration over the true natural signal. This is due to the fact that the low-pass filter conserves the zeroth Fourier mode of the true natural signal. For this reason it is also not possible to fix the negative contributions by cutting them away or taking the absolute values of the low-pass filtered signal, as this operations will not conserve the zeroth Fourier mode and therefore violates conserved quantities like the total number or total mass of galaxies.

\section{Sampling 3d galaxy distributions}
\label{Sampling_3d_galaxy_distributions}
In the sections above we examined the theory of discretizing continues signals and demonstrated some of the problems coming with this approach. For simplicity we discussed only the 1d case, but all results can be extended straight forwardly to the 3d case.
In this section we will discuss the practical implementation of sampling methods in order to process galaxy distributions via FFT techniques.

\subsection{Ideal sampling procedure}
\label{Sampling_3d_galaxy_distributions}
As already described in section \ref{Instrument_Response} a galaxy distribution can be expressed as a set of point particles. With this definition a 3d galaxy distribution consisting of \(M\) galaxies can be expressed as:
\begin{equation}
\label{eqn:Gal_dista}
\rho(\vect{x})= \sum_{p=1}^{M} \delta^D(\vect{x}-\vect{x_p}) \, ,
\end{equation}
where \(\vect{x_p}\) is the three dimensional position vector of a galaxy.
\begin{figure*}
	\centering
	{
	\begin{picture}(100,180)

\put(-180,180){a)}

  	\put(-180,180){
	\rotatebox{270}
	{
		\includegraphics[bb = 61 174 538 646,width=0.35\textwidth,clip=true]{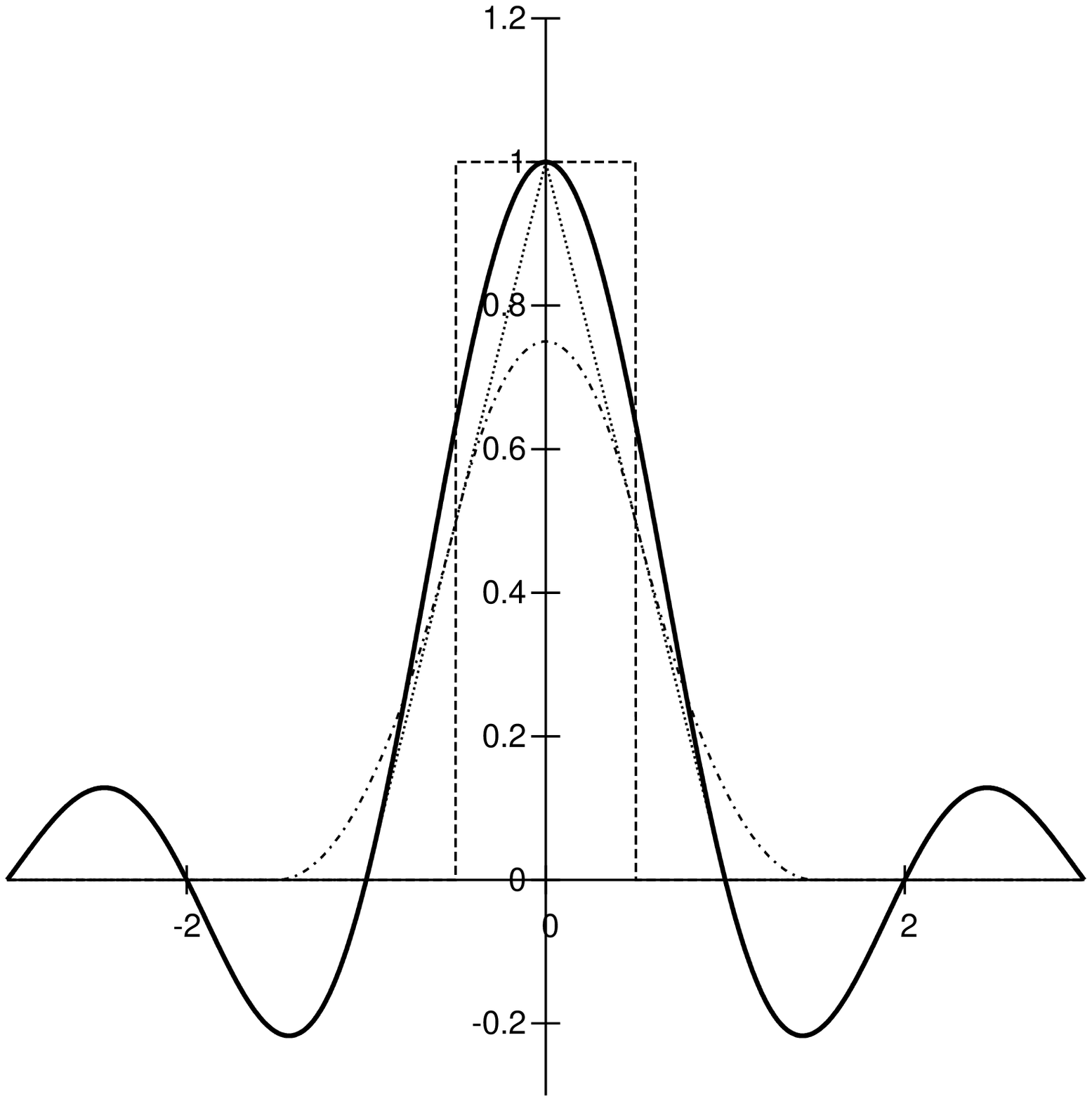}
	}
	}

\put(80,180){b)}
	
	\put(80,180){
	\rotatebox{270}
	{
		\includegraphics[bb = 61 174 538 646,width=0.35\textwidth,clip=true]{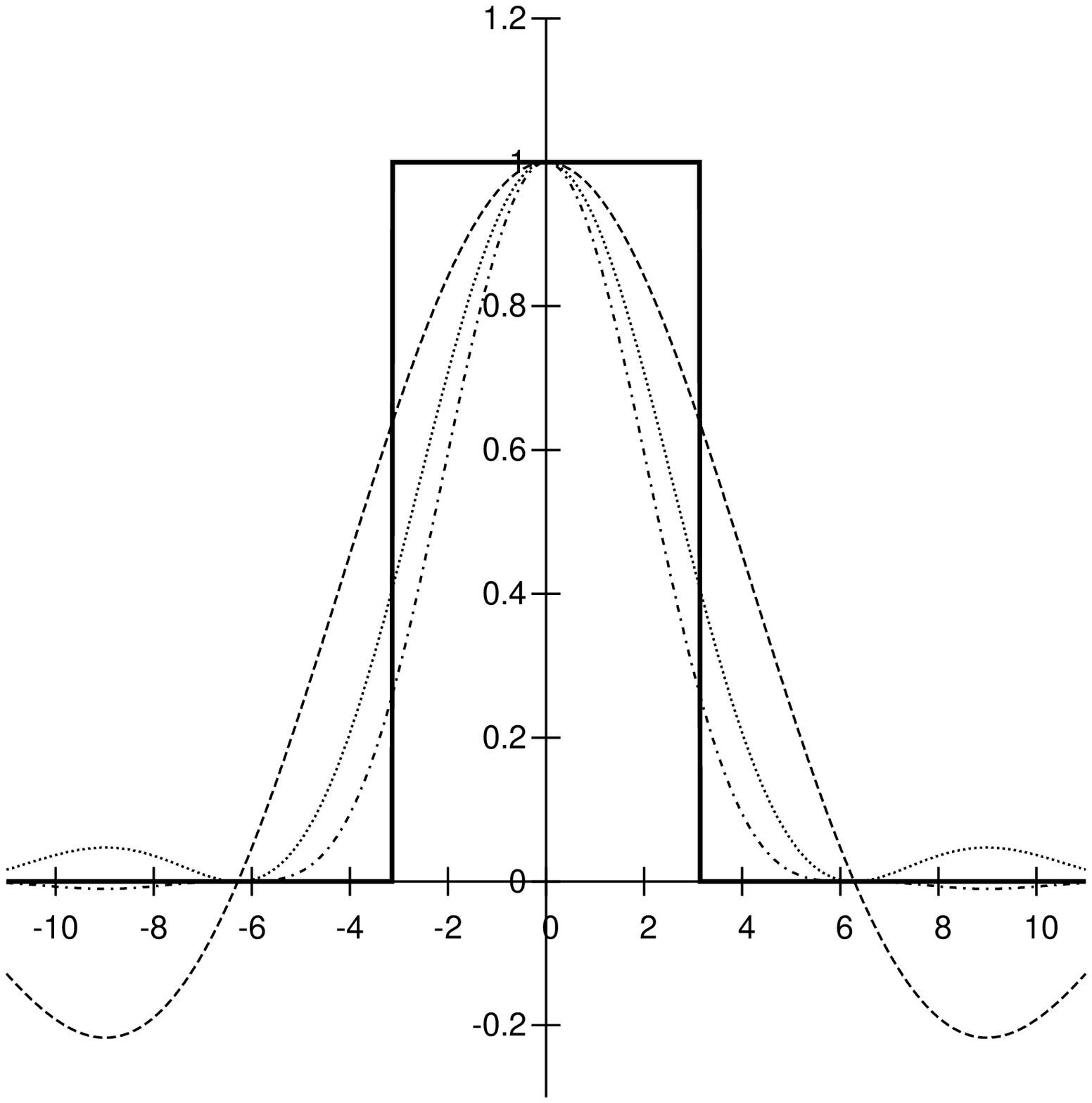}
	}
	}

        \end{picture}
	}
	\caption{Comparison of the real (a)  and Fourier-space (b) representations of the NGP kernel (dashed curve), the CIC kernel (dotted curve) and the TSC kernel (dashed dotted curve) to the ideal low-pass filter (solid curve).}
	\label{fig:FILTER_FT}
\end{figure*}
Note, that assuming a galaxy to be a point particle is a strong prior. Therefore in more general it is possible to allow for the galaxies to have a shape \(s(\vect{x})\), which describes the physical extend of a galaxy into space. The new density field \(\rho'(\vect{x})\) is then obtained by simple convolution of equation (\ref{eqn:Gal_dista}) with the galaxy shape \(s(\vect{x})\):
\begin{equation}
\label{eqn:Gal_distb}
\rho'(\vect{x})= \left(s\circ \rho\right)(\vect{x}) = \sum_{p=1}^{M} s(\vect{x}-\vect{x_p}) \, ,
\end{equation}
However, such a modification of the galaxy shape would only be necessary if the grid resolution is sufficient enough to resolve individual galaxies and their inner structure, which is not the usual case for applications as discussed in this work.
Therefore, in the following a galaxy distribution will always be considered to be a set of point particles.

The galaxy distribution as given by equation (\ref{eqn:Gal_dista}) can be sampled to a grid according to the ideal sampling scheme displayed in Figure \ref{fig:sampling_scheme}, which consists of two steps.
In the first step the high frequency contributions of the galaxy distribution are eliminated by applying the ideal low-pass filter:
\begin{equation}
\label{eqn:Gal_dista_filtered}
\varrho(\vect{x})= \left(W\, \circ \rho\right)(\vect{x})=\sum_{p=1}^{M} W(\vect{x}-\vect{x_p}) \, ,
\end{equation}
which simply states that the galaxies have been given the artificial shape of the 3d sinc function \(W(\vect{x})\). In a following step the low-pass filtered distribution will be sampled at discrete positions on the grid. For practical implementation these two steps will usually be done in one step, by simply convolving the galaxy distribution to the discrete gridpoints:
\begin{equation}
\label{eqn:Gal_dista_sampled}
\varrho_{\vect{n}}=\sum_{p=1}^{M} W(\Delta x\, \vect{n}-\vect{x_p}) \, ,
\end{equation}
where \(\Delta x\) is the grid spacing and \(\vect{n}\) is the triplet of integers \(\vect{n}=i,j,k\). The sum in equation (\ref{eqn:Gal_dista_sampled}) describes the ideal sampling procedure, which allows to have the best information conservative representation of the continuous signal in discrete form. However, as the ideal low-pass filter kernel extends over all space, the sum over all particles has to be evaluated for each individual voxel of the discrete grid, which makes this procedure impractical for real world applications.

\section{Practical sampling}
\label{practical_sampling}
Due to the infinite support of the ideal low-pass filter kernel in real-space the ideal sampling method (by convolving with such a filter) is in general not feasible. For many applications it is computationally too expensive. For this reason one usually chooses a practical approach by approximating the ideal sampling operator, by a less accurate, but faster calculable sampling operator. This usually means to approximate the low pass filter, by a function with compact support in real-space. As a result the convolution can be calculated faster, for the prize of not completely suppressing the aliasing power in the stop-bands.
Approximating the ideal sampling operator therefore is always a trade off between accuracy and computational speed.

In the literature of Digital Signal Processing this problem is well known and studied for many years already. Especially the literature of modern 3D computer graphics provides a lot of practical solutions to the problem of sampling and filtering. Many very detailed studies about the optimal filter approximations have been made, and can be found in the signal processing literature \citep{Hauser2000,Mitchell1988,Marschner1994,Wolberg1997}.

In the following we will discuss some sampling techniques as commonly used in cosmology and display their strength and weaknesses.

\subsection{Filter Approximation}
\label{Filter_Approximation}
There are several known practical issues when dealing with filter approximation \citep{Marschner1994}.
Using filter or filter approximation will  in general lead to a variety of effects like smoothing ("blurring"), which refers to the removal of rapid variations in the signal by spatial averaging. Or, one will observe a reshaping of the Fourier-space by applying a filter which has not unity gain at every mode in Fourier-space. In addition, low-pass filtering step discontinuities will result in oscillations or ringing, just before or after discontinuities \citep{Marschner1994}. This is the Gibbs phenomenon, and is due to the fact, that step discontinuities in the signal cannot be represent by a finite superpostion of Fourier waves. 

In cosmology usually the sub-class of separable low-pass filter approximations are used \citep{HOCKNEYEASTWOOD1988}. These filter obey the relation \citep{Marschner1994}:
\begin{equation}
\label{eqn:Seperable_filter}
W(x,y,z)=W_s(x)\, W_s(y)\, W_s(z) \, ,
\end{equation}
where \(W_s(x)\) is the separated 1D filter kernel. This separation allows for fast computational speed in performing the 3D convolution \citep{Ferguson2001}.

The simplest filter approximation frequently used in cosmology is the Nearest Grid Point (NGP) kernel. 
This zero-order kernel provides the simplest and fastest interpolation method \citep{Duan2003}. Each galaxy in the input data is assigned to the value of the nearest sample point in the output data. The nearest neighbor kernel is defined as:
\begin{eqnarray}
\label{eqn:NGP}
W_s(x) = \left \{ \begin{array}{ll}
  1 & \quad \mbox{for $-0.5< x \le 0.5 $}\\
  0 & \quad \mbox{otherwise}\\ \end{array} \right.
\end{eqnarray}

The Fourier response of this kernel is a sinc function which has a poor localization and pass-band selectivity. This property typically leads to low-quality interpolated data with blocking effects for signals with high frequency contents like sharp intensity variations or high noise level as might be expected in sampling individual point particles \citep{Duan2003}. 

Another commonly used filter is the linear filter. This first-order kernel linearly interpolates between adjacent points of the input data along each dimension \citep{Duan2003}. It is defined as:
\begin{eqnarray}
\label{eqn:CIC}
W_s(x) = \left \{ \begin{array}{ll}
  1-|x| & \quad \mbox{for $ 0\le |x| <1 $}\\
  0 & \quad \mbox{for $ 1\le |x|$}\\ \end{array} \right.
\end{eqnarray}
In Cosmology this filter is often referred to as the Cloud in Cell (CIC) scheme \citep{HOCKNEYEASTWOOD1988}. It is popular for prefiltering as it provides a good tradeoff between filtering quality and computational cost \citep{Duan2003}. Nevertheless, a significant amount of spurious stop-band aliasing components continues to leak into the pass-band, contributing to some aliasing \citep{Wolberg1997}. 

The next order kernel is the so called Triangular Shaped Cloud (TSC) kernel which is defined as:
\begin{eqnarray}
\label{eqn:TSC}
W_s(x) = \left \{ \begin{array}{ll}
  0.75-|x|^2 & \quad \mbox{for $ 0\le |x| <0.5 $}\\
  0.5\,(1.5-|x|)^2 & \quad \mbox{for $ 0.5\le |x| <1.5 $}\\
  0 & \quad \mbox{for $ 1.5\le |x| $} \end{array} \right. \, ,
\end{eqnarray}
\citep{HOCKNEYEASTWOOD1988}.
This filter kernel has better stop-band behaviour and is therefore usually preferred to the CIC kernel.
 
This can be seen in Figure \ref{fig:FILTER_FT} where the Fourier transforms of the NGP, CIC and the TSC filter kernel are compared to the ideal low-pass filter. In addition to the stop-band leakage of all filters they all introduce pass-band attenuation, meaning the power of the Fourier modes is suppressed by the filter. This is of special relevance in all applications where one is interested in the correct spectral representation of the signal, in particular in the case of the matter power spectrum estimation.
\begin{figure*}
	\centering
	{
	\begin{picture}(100,160)

\put(-200,80){\rotatebox{90}{P(k)}}
\put(-30,80){\rotatebox{90}{P(k)}}
\put(140,80){\rotatebox{90}{P(k)}}
\put(-130,0){k [h/Mpc]}
\put(40,0){k [h/Mpc]}
\put(210,0){k [h/Mpc]}

\put(-200,160){a)}

  	\put(-200,160){
	\rotatebox{270}
	{
		\includegraphics[bb = 61 159 546 666,width=0.3\textwidth,clip=true]{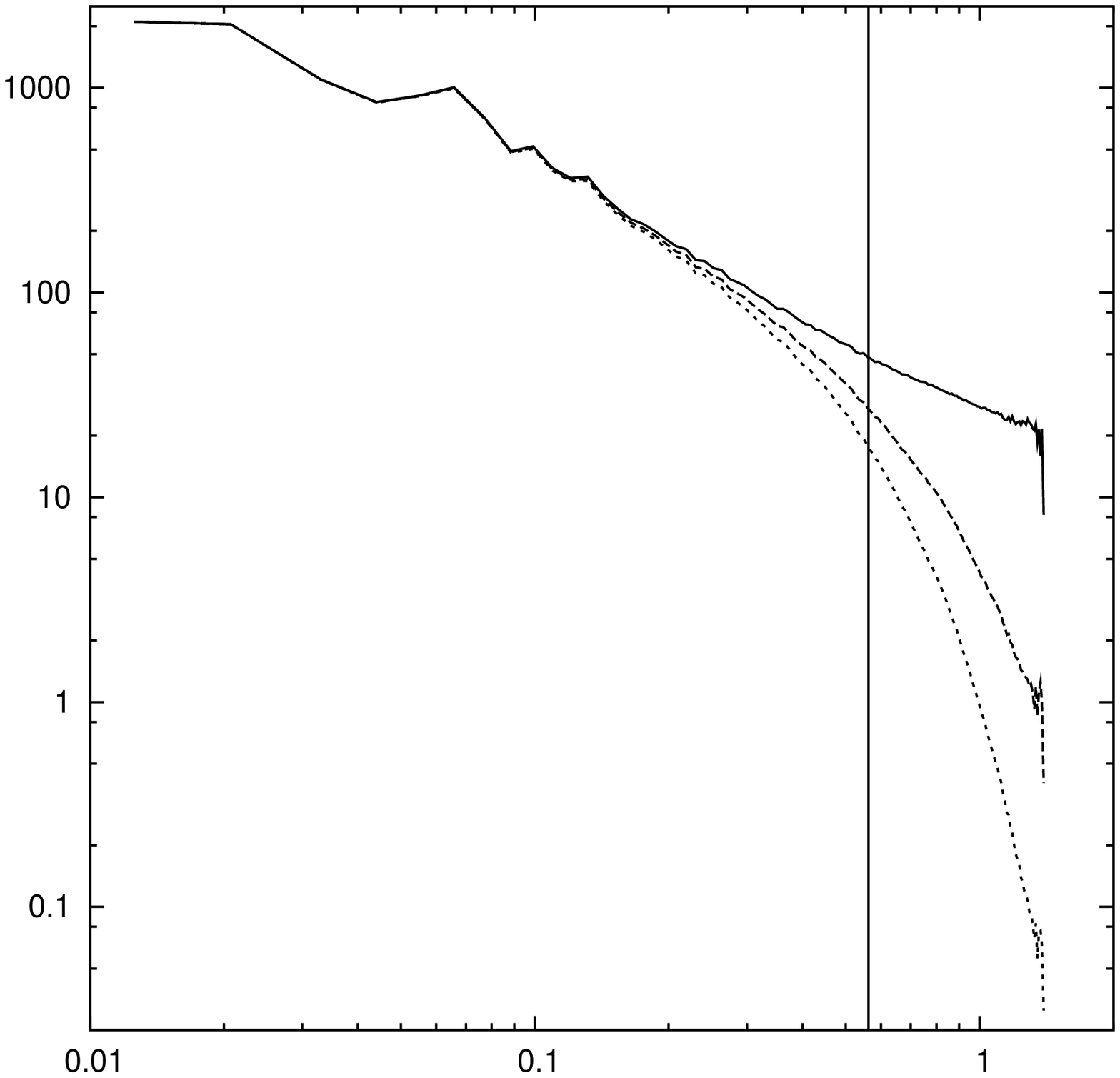}
	}
	}

\put(-30,160){b)}
	
	\put(-30,160){
	\rotatebox{270}
	{
		\includegraphics[bb = 61 159 546 666,width=0.3\textwidth,clip=true]{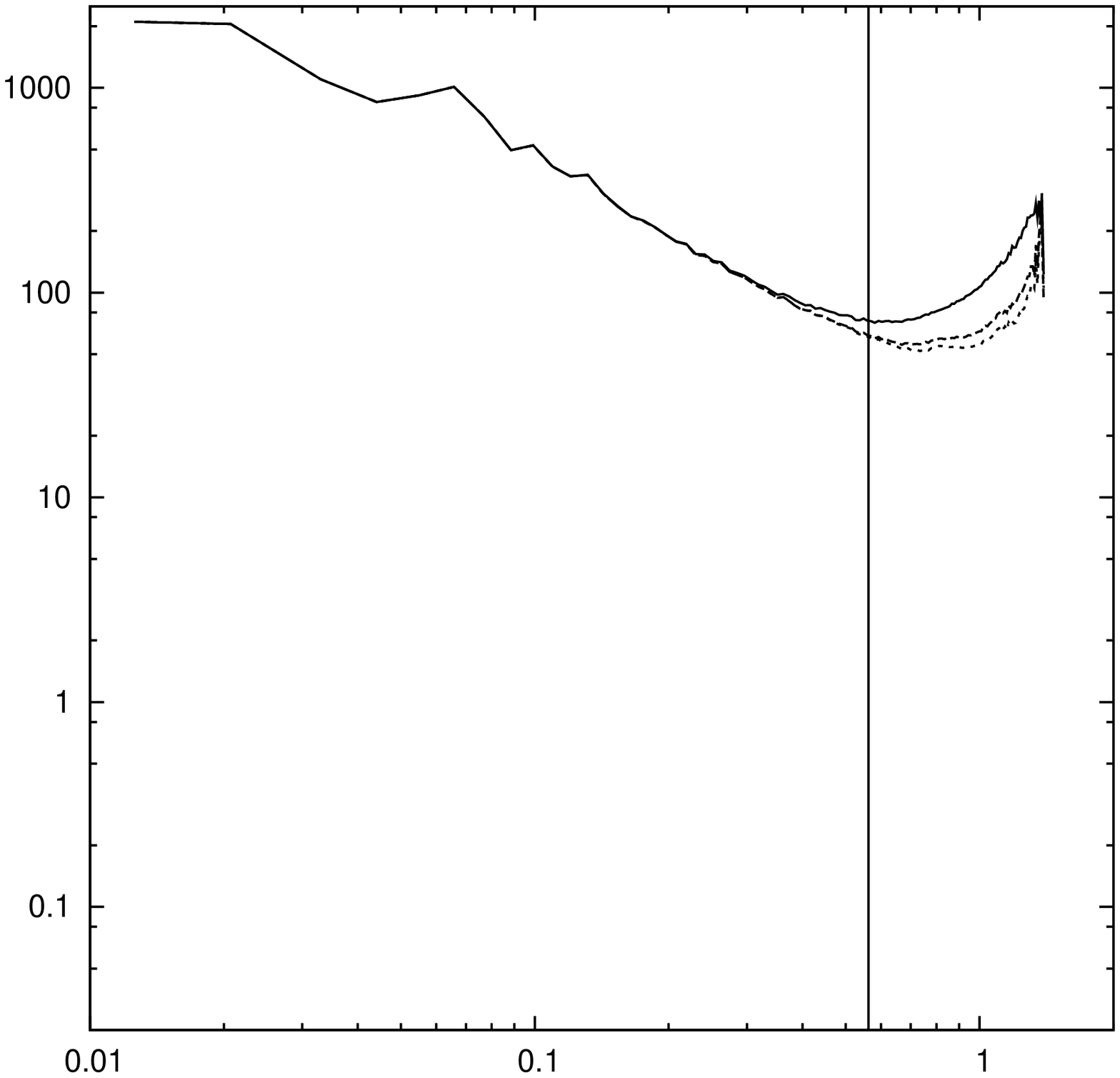}
	}
	}

\put(140,160){c)}
	
	\put(140,160){
	\rotatebox{270}
	{
		\includegraphics[bb = 61 159 546 666,width=0.3\textwidth,clip=true]{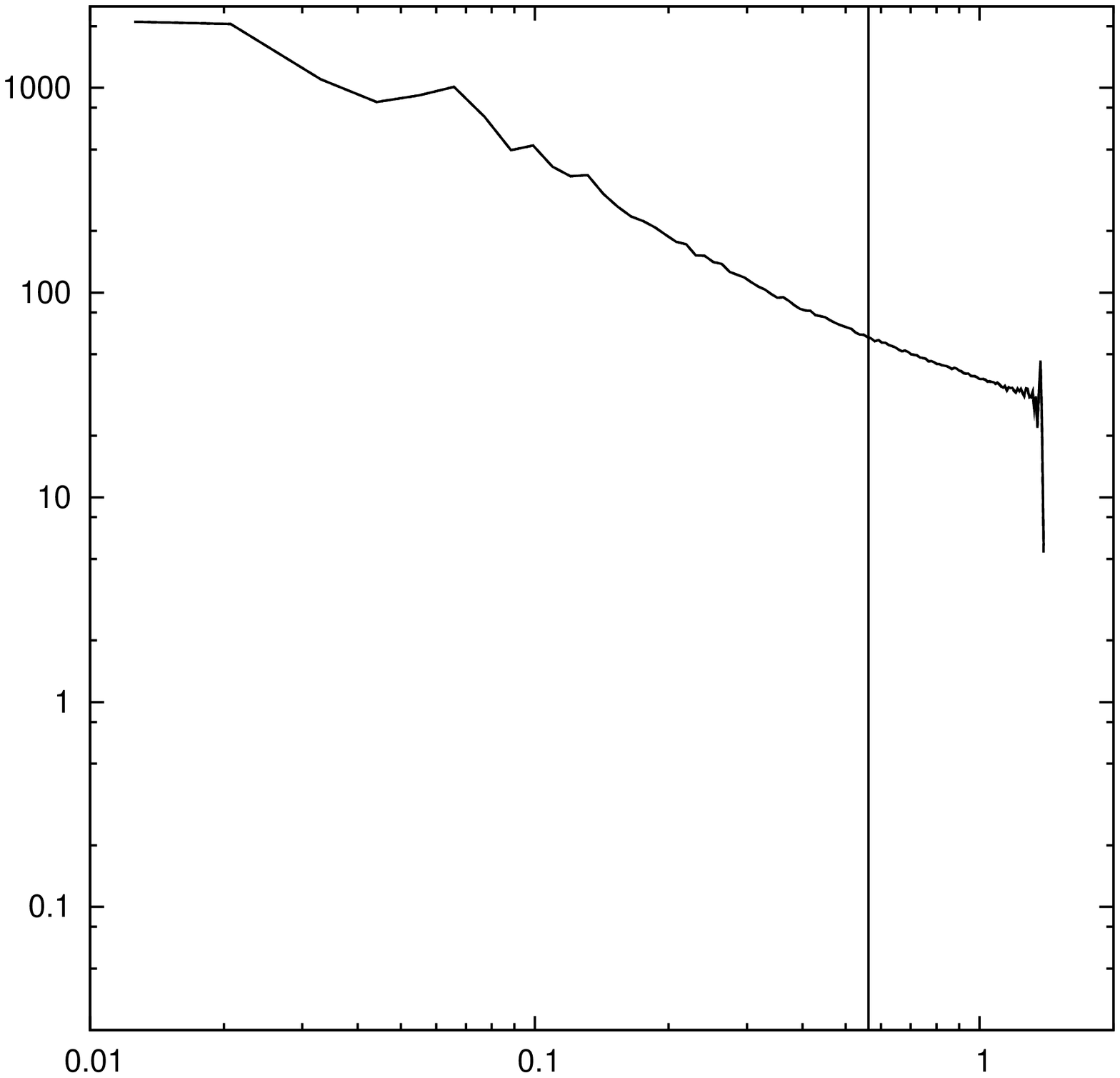}
	}
	}

        \end{picture}
	}
	\caption{Power spectra calculated from a galaxy mock catalog created by \citet{2007MNRAS.375....2D} with the three different filter methods NGP (solid line), CIC (long-dashed line) and TSC (short-dashed line). Panel a) displays the power spectrum obtained without further processing, while panel b) demonstrates the effect of deconvolving with the filter kernel. In panel c) we show the power spectrum taken using the supersampling technique, prefilterd with CIC kernel (solid line) and TSC kernel (dashed line) which cannot be distinguished visually. The vertical line indicates the critical Fourier mode \(k_c\) below which the spectrum usually is trusted.}
	\label{fig:PWSPEC_different_kernel}
\end{figure*}

This reshaping of the Fourier-space due to the application of an imperfect low-pass filter can be undone by deconvolving with the Fourier transform of the corresponding filter kernel. This procedure in fact returns the correct power at most Fourier modes, as can be seen in Figure \ref{fig:PWSPEC_different_kernel}, where we demonstrate the power spectra obtained after applying these filter approximations to a galaxy mock catalog created by \citet{2007MNRAS.375....2D}. The left panel displays the power spectrum without deconvolving by the filter kernel while the right shows the same power spectra but deconvolved by the corresponding Fourier transforms of the filter kernels. As it can be seen in Figure \ref{fig:PWSPEC_different_kernel} most of the modes now seem to have correct power, and the different filter approximations agree to each other at the lower modes. Nevertheless, this procedure will also enhance the aliased contribution leaking into the pass-band region, which will severely affect the power spectrum towards the highest Fourier modes.
In general all Fourier modes of the sampled signal will be affected by aliasing. However, natural signal Fourier amplitudes tend to drop off with higher frequency and therefore the sampled signal Fourier modes will be affected less and less towards the lower frequencies.

For this reason the power spectrum is usually trusted only up to Fourier modes of \(k_{c}=0.7 \pi\, N /L\) \citep{Volker2008},in cosmology.
Figure \ref{fig:PWSPEC_different_kernel} also demonstrates that the NGP method performs poorly and should not be considered to be used in accuracy applications.

Another important thing to remark is that although TSC couples \(3^3\) cells, while CIC only couples only \(2^3\), and therefore has a broader spatial support, it does not perform much better than the CIC scheme. This is due to the fact, that the overall amplitude of the ideal low-pass filter drops off in space, and hence the main contribution of the corrections will come from cells close to the origin. As the amplitude of the ideal low-pass filter decreases rapidly it requires more and more cells to make as strong corrections as were obtained by only considering the closer cells to the kernel origin.
This demonstrates that in order to make further corrections to the filter approximation, one requires to couple more and more cells, while at the same time corrections will be smaller and smaller.

\section{Supersampling}
\label{Supersampling}
As already pointed out above, correct sampling is in practice generally not feasible, and some approximation must necessarily be made. However, for most natural signals, like the galaxy distribution, an approximation method can take advantage of the fact that signal content generally declines as the frequency increases. This observation resulted in the rule of thumb, that the matter power spectrum can be trusted at frequencies below \(k_c\) \citep{Volker2008}. The most common anti-aliasing technique in 3D computer graphics, supersampling, takes advantage of this by sampling the signal at a frequency higher than the desired sample rate \citep{GOSS2000}. A low-pass filter is then applied to the supersamples which attenuates or eliminates the frequency content above a threshold so that the signal resampled at the target rate exhibits fewer aliasing artifacts.

As supersampling methods are very successful in reducing sampling artifacts in 3D computer graphics, we present an adapted supersampling method to be used in cosmological applications.

\subsection{Super resolution and downsampling}
The supersampling method consist of two main steps, the supersampling step, in which the signal is sampled at high resolution, and the downsampling process, in which the high resolution samples are sampled to the target resolution. In our approach we make use of FFTs to allow for pass-band attenuation correction and for fast and efficient calculation of the overall supersampling method.
This two stage filtering process is illustrated in Figure \ref{fig:sampling_scheme}, and consists of the following steps:
\begin{enumerate}
\item{supersampling:
The continuous signal is sampled to a grid with a resolution \(n\) times larger than the target resolution. This is achieved by applying the CIC or TSC method to allow for fast and efficient computation of the high resolution samples.}
\item{downsampling:
The high resolution samples are corrected for pass-band attenuation and low-pass filtered. The high resolution low-pass filtered samples are then  resampled at the target resolution.}
\end{enumerate}

While the supersampling step is in principle identical to the method as described in section \ref{practical_sampling} the downsampling process requires some more explanation.

As already described in section \ref{practical_sampling}, using an imperfect filter approximation usually leads to pass-band attenuation which has to be corrected.

This is achieved by applying a FFT to the supersampled signal, and then dividing by the Fourier transform of the according filter approximation. Reducing the supersampled signal to the final resolution requires to low-pass filter the supersamples and sample them again at lower resolution.

In our approach low-pass filtering and downsampling is done in one step. The low-pass filter step can easily be achieved in Fourier-space by discarding all frequencies higher than a given threshold. 
Introducing the ideal low pass-filter in discrete Fourier-space as:
\begin{eqnarray}
\label{eqn:discrete_ideal_LP}
\hat{W}_k = \left \{ \begin{array}{ll}
  1 & \quad \mbox{for $ -\left (\frac{N_{SS}}{2n}-1 \right )\le k \le \frac{N_{SS}}{2n} $}\\
  0 & \quad \mbox{otherwise}\\ \end{array} \right. \, ,
\end{eqnarray}
with \(N_{SS}\) being the number of supersampling cells, yields the low-pass filtered supersampled signal \(g_j\) as:
\begin{eqnarray}
\label{eqn:discrete_real_SS_signal}
g_j &=& C_{SS}\sum_{k=-\left(\frac{N_{SS}}{2}-1\right)}^{\frac{N_{SS}}{2}}\hat{W}_k \, \hat{f}_k\, e^{2\pi j k \frac{\sqrt{-1}}{N_{SS}}} \nonumber \\ 
&=& C_{SS}\sum_{k=-\left(\frac{N_{SS}}{2\,n}-1\right)}^{\frac{N_{SS}}{2\,n}} \, \hat{f}_k\, e^{2\pi j k \frac{\sqrt{-1}}{N_{SS}}} \, , 
\end{eqnarray}
where \(\hat{f}_k\) is the pass-band attenuation corrected Fourier transform of the supersampled signal and \(C_{SS}\) is the FFT normalization constant according to the super resolution Fourier transform. Downsampling can now easily be achieved by introducing the number of cells of the target resolution \(N_{DS}=N_{SS}/n\) and using \(j=n\,j'\), which simply results from the fact that the grid spacing of the target resolution grid is \(n\) times smaller than the super resolution grid spacing.
\begin{eqnarray}
\label{eqn:discrete_real_SS_signal_a}
g_j &=& C_{SS}\sum_{k=-\left(\frac{N_{DS}}{2}-1\right)}^{\frac{N_{DS}}{2}} \, \hat{f}_k\, e^{2\pi j' k \frac{\sqrt{-1}}{N_{DS}}} \, . 
\end{eqnarray}
Sampling the supersampled and low-pass filtered signal to the target resolution grid can be done by applying the Kronecker delta function \(\delta^k_{jj'}\) to yield the downsampled signal: 
\begin{eqnarray}
\label{eqn:discrete_real_DS_signal}
g_{j'} &=& C_{SS}\sum_{k=-\left(\frac{N_{DS}}{2}-1\right)}^{\frac{N_{DS}}{2}} \, \hat{f}_k\, e^{2\pi j' k \frac{\sqrt{-1}}{N_{DS}}}\nonumber \\
&=& C_{DS}\sum_{k=-\left(\frac{N_{DS}}{2}-1\right)}^{\frac{N_{DS}}{2}} \,\frac{C_{SS}}{C_{DS}}\, \hat{f}_k\, e^{2\pi j' k \frac{\sqrt{-1}}{N_{DS}}} \, , 
\end{eqnarray}
with \(C_{DS}\) being the FFT normalization according to the target resolution.

Hence, equation (\ref{eqn:discrete_real_DS_signal}) provides an easy prescription for the low-pass filtering and downsampling process. Downsampling can be achieved by sorting the Fourier modes of the supersampled signal up to the Nyquist mode of the target resolution in the target resolution grid in Fourier-space, rescale it to the correct Fourier normalization and apply an inverse FFT to obtain the real-space representation for the signal. 

The interpretation of this process is easy. As the FFT assumes the signal to be periodic and to consist of a finite superposition of Fourier waves, the resolution of the supersampled signal can simply be reduced by discarding the high frequency waves, resulting in a smoother real-space representation of the sampled signal.
The result of this procedure is a sampled signal at target resolution which is greatly cleaned from aliasing effects.

The ability of this supersampling method is demonstrated with Figure \ref{fig:PWSPEC_different_kernel}. Here we used a super resolution factor \(n=2\) and calculated the power spectrum from the supersampled signal. For the supersampling step we applied the CIC and the TSC method. As can be seen in Figure \ref{fig:PWSPEC_different_kernel} the two results for CIC and TSC cannot be distinguished. Also it is worth noticing that the aliasing contribution had been corrected far beyond the usual limit of trust \(k_{c}=0.7 \pi\, N /L\).

\begin{figure*}
	\centering
		\rotatebox{270}
	{
		\includegraphics[bb = 0 0 334 615,width=0.55\textwidth,clip=true]{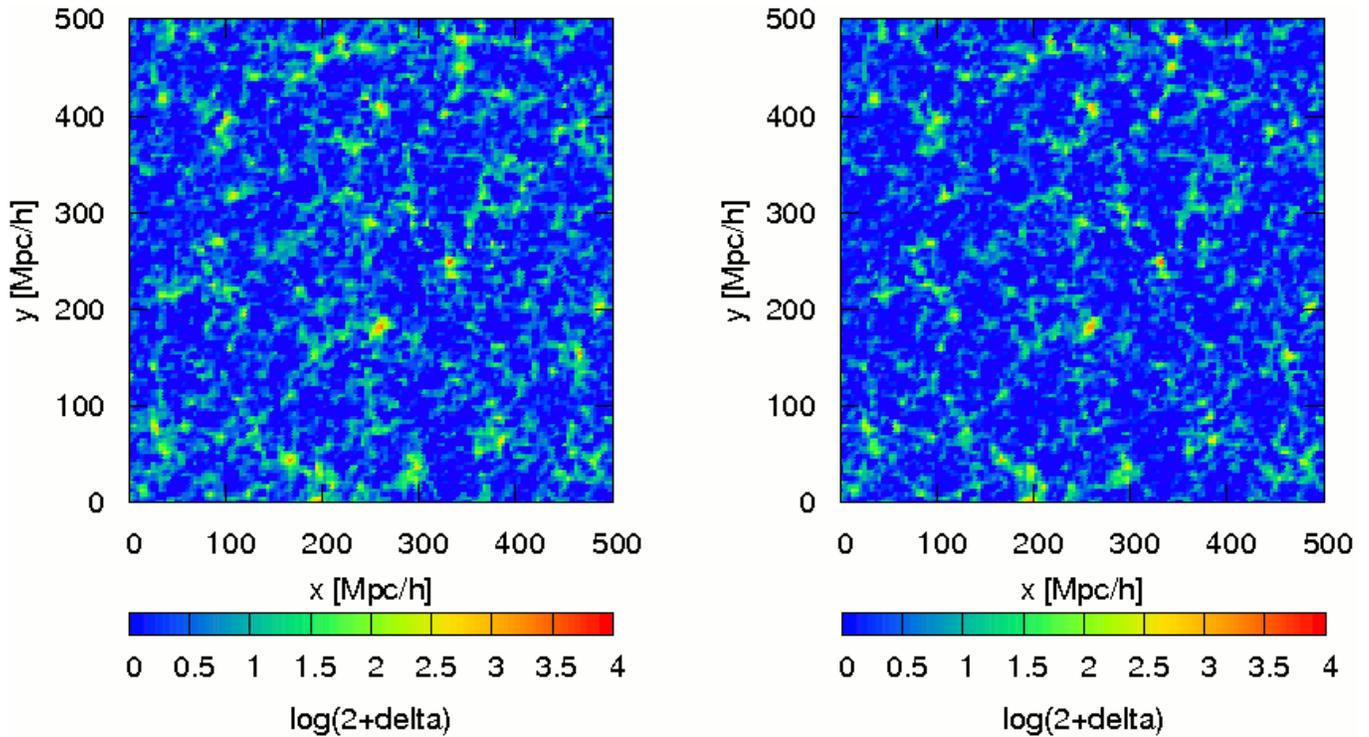}
	}
	\caption{Comparison of slices through sampled mock galaxy distribution using the supersampling method (left panel) and the standard TSC method (right panel).}
	\label{fig:slices_SS_TSC}
\end{figure*}

In Figure \ref{fig:slices_SS_TSC} we contrast the supersampling method to the TSC method by comparing slices through a sampled 3D mock galaxy catalogue in a \(500\unit{Mpc}\) box which has been sampled regularly at \(128^3\) pixels.
At first glance, visually there is not much difference between the two samples. However, a closer inspection of the high density regions reveals that the density field obtained with the pure TSC method is more blocky, meaning it can have sudden density jumps from one pixel to the next. In the case of the supersampled density field these high density pixels are surrounded by pixels with varying densities. In this way the sharp edges are softened, as should be expected by an anti-aliasing technique. Additionally, one can observe, that the filaments are better connected in the case of the supersampled density field. This is due to the additional super resolution information which was taken into account in the supersampling process.

\section{Discussion and Conclusion}
\label{Discussion}
In this work we reviewed and discussed the use of FFT techniques in cosmological signal processing procedures. As described in section \ref{The_requirements_of_FFTs} the application of FFTs requires the signal to be discrete in real and Fourier-space. As natural signals are in general continuous in either spaces, they must be approximated by discretized representations with the constrained of conserving as much information of the true natural signal as possible.
According to Shannon's theorem this can be achieved by low-pass filtering and sampling the continuous signal at discrete positions.

However, we also demonstrated that Shannon's theorem is a necessary but not sufficient requirement, for processing a signal via FFTs or in more general DFTs. The application of FFTs or DFTs additionally requires the signal to be periodic on a bounded spatial domain, or in other words, the signal must be decomposable into a finite amount of Fourier waves. Thus, the application of FFT or DFT techniques implicitely assumes the discreteness of Fourier-space which is generally not given by natural signals. This introduces an additional filtering procedure for the true continuous Fourier modes, as shown in section \ref{FFTs_and_the_Fourier_space_representation}. 

In section \ref{Instrument_Response} we demonstrated that, due to discretizing, the physical property of being a positive density field is not reflected by the discretized representation of the continuous galaxy distribution. This is due to the convolution with the ideal low-pass filter kernel which is a sinus cardinal function and thereby not positive definite.
The loss of physicality is thus an expression of the loss of information due to discretizing the continuous signal, and hence is a fundamental problem of the method itself. 

In addition to this fundamental problems, we discussed that ideal sampling is in general not applicable to real world problems. 

Signal processing in cosmology or any other field is a technological challenging problem governed by the requirement of computational feasibility of the sampling procedure. As ideal sampling is in general computational too expensive to be used for real world signal processing applications one usually introduces low-pass filter approximations to allow for faster computation for the expense of introducing sampling artifacts like aliasing.

It is, however, often exactly this leakage of aliasing power from the stop-band which makes further processing of the sampled signal difficult. For example estimations of higher-order spectra with FFT techniques will be erroneous due to sampling artifacts. As mentioned in \citet{Volker2008}, so far there is no known approach to accurately correct for the sampling effects in measuring higher-order spectra like the bi-spectrum with FFT techniques. Also note, that iterative signal reconstruction processes, which utilize operations like real-space multiplications, deconvolutions and other signal processing operations, may enhance or distribute this false aliased power into regions of the power spectrum formerly unaffected by aliasing.

The aim for signal processing technologies therefore is to find a low-pass filter approximation which sufficiently suppresses these sampling artifacts, while at the same time still being computationally far less expensive than the ideal sampling procedure. The Digital Signal Processing literature provides plenty of possible approaches (see e.g. \citet{DSPGUIDE2002}) which could be introduced to various cosmological signal processing problems.

Recently many authors presented new methods to reduce aliasing effects in cosmological power spectrum estimation due to imperfect filter approximation.
\citet{Jing2005} proposes to evaluate all aliasing sum, and additionally uses an iterative scheme, which requires a priori knowledge about the true power spectrum, to correct for the aliasing effects. More recently \citet{Volker2008} demonstrated that by using scale functions of the Daubechies wavelet transformation as an approximation to the ideal low-pass filter aliasing can be greately suppressed. This is achieved by allowing the filter approximation to extend over a larger support in real-space.

Enlarging the spatial support of the low-pass filter kernel will lead to a better approximation of the ideal low-pass filter. Hence, by approximating the ideal sinc function closer and closer aliasing can be suppressed below any limit.

This approach is well known in the DSP literature, which favors the windowed sinc functions as optimal approximations to the ideal low pass filter \citep{Hauser2000,Duan2003,Marschner1994,DSPGUIDE2002}.

As it tends towards evaluating the sum given in equation (\ref{eqn:Gal_dista_sampled}), this approach of using filter with greater and greater spatial 
support becomes more and more computationally impractical.

For these reasons we introduced a new supersampling technique frequently applied as an anti-aliasing technique in 3D computer graphics \citep{Wolberg1997,GOSS2000}.

As described in section \ref{SAMPLING_THEOREM}, aliasing is a result of sampling, or more specifically, the lack of a sufficient sampling rate. 

Supersampling, as the name suggests, solves the aliasing problem by taking more samples than would normally be the case with usual particle assignment schemes as CIC or TSC. By taking more samples at sub-pixels, we are able to more accurately capture the details of the natural continuous signal. The target pixel value is then obtained by averaging the values of the sub-pixel reducing the aliasing edge effects in the signal.
Hence, supersampling reduces aliasing by bandlimiting the input signal and exploiting the fact that the signal content generally declines as the frequency increases.

The supersampled signal still contains artifacts due to aliasing, but the artifacts are less prominent as if the signal was sampled directly at the target rate.
Nevertheless, there are two problems associated with supersampling. The first problem is, as already mentioned, that the newly designated Nyquist frequency of the superresolution samples continues to be fixed \citep{Wolberg1997}. Hence, there will always be sufficiently higher frequencies that will alias. The second problem is cost of memory \citep{Wolberg1997}. To sample a 3D signal at double resolution requires eight times more memory. On the otherhand, if memory is no issue, the supersampling method is computationally less expensive than introducing a next order low-pass filter approximation, while at the same time providing better aliasing suppression. In addition, the supersampling procedure, as proposed in this paper, incorporates pass-band attenuation corrections, which otherwise would have to be corrected for in an additional separate step, when applying just a low-pass filter approximation.

This supersampling technique, therefore, can be understood as being complementary to the approach of finding better low-pass filter approximations. The combination of better filter approximations with the supersampling method will naturally increase the quality of the sampled signal further.

Nevertheless, the problem of signal processing in cosmology via FFT techniques is a complex one. Beside being subject to fundamental issues of discritizing physical quantities, it mainly is a technological challenging problem, which requires to efficiently apply a low-pass filter to an observed continuous signal. The optimal solution for each individual application may also vary from case to case, as it can be dependent on the individual properties of the underlying signal and scientific application (e.g high accuracy power spectrum estimation). 

We therefore believe, that the supersampling method presented in this paper will be useful for cosmological signal processing, permitting highly accurate power spectrum estimates.

\section*{Acknowledgments}
We thank the "Transregional Collaborative Research Centre TRR 33 - The Dark Universe" and the Marie Curie FP7 fellowship for the support of this work.

\bibliography{paper}
\bibliographystyle{mn2e}

\appendix
\section{Continuous Fourier transformation}
We define the Fourier transformation as:
\begin{equation}
\label{eqn:CFT}
\hat{f}(k)=\int_{-\infty}^{\infty} f(x)e^{-ikx} dx \,
\end{equation}
and the according inverse Fourier transform is defined as:
\begin{equation}
\label{eqn:ICFT}
f(x)=\frac{1}{2\pi} \int_{-\infty}^{\infty} \hat{f}(k)e^{ikx} dk \,
\end{equation}

\subsection{Convolution Theorem}
\label{convolution_theorem}
The convolution theorem states that a convolution in real-space is a product in Fourier-space, and a convolution in Fourier-space is a product in real-space.
This can be shown as follows.

Let \(h(x)\) be the convolution of \(f(x)\) with \(g(x)\) defined as:
\begin{equation}
\label{eqn:CFT_Convolution}
h(x)=\int_{-\infty}^{\infty} f(y)\,g(x-y) dy \, .
\end{equation}
Applying the Fourier transform  \ref{eqn:CFT} to \(h(x)\) yields:
\begin{eqnarray}
\label{eqn:Convolution_theo_a}
\hat{h}(k) &=& \int_{-\infty}^{\infty} h(x)e^{-ikx} dx \nonumber \\
&=& \int_{-\infty}^{\infty} f(y) \int_{-\infty}^{\infty} e^{-ikx}\,g(x-y) dy dx \nonumber \\
&=& \int_{-\infty}^{\infty} e^{-iky} f(y) dy \int_{-\infty}^{\infty} e^{-iky'}\,g(y') dy'  \nonumber \\
&=& \hat{f}(k)\,\hat{g}(k) \, ,
\end{eqnarray} 
where we substitute with \(y'=x-y\) and used \ref{eqn:CFT} to write the last line.
Hence, a convolution of the functions \(f(x)\) and \(g(x)\) in real-space is a product of the respective Fourier transforms \(\hat{f}(k)\) and \(\hat{g}(k)\) in Fourier-space.

The second part of the prove is analog to the first one presented above.

Let \(\hat{h}(k)\) be the convolution of \(\hat{f}(k)\) with \(\hat{g}(k)\) defined as:
\begin{equation}
\label{eqn:ICFT_Convolution}
\hat{h}(k)=\frac{1}{2\, \pi}\int_{-\infty}^{\infty} \hat{f}(p)\,\hat{g}(k-p) dp \, .
\end{equation}
Applying the inverse Fourier transform  \ref{eqn:ICFT} to \(\hat{h}(k)\) yields:
\begin{eqnarray}
\label{eqn:Convolution_theo_b}
h(x) &=& \frac{1}{2\, \pi} \int_{-\infty}^{\infty} \hat{h}(k)\,e^{ikx} dk \nonumber \\
&=& \frac{1}{(2\, \pi)^2} \int_{-\infty}^{\infty} \hat{f}(p) \int_{-\infty}^{\infty} e^{ikx}\,\hat{g}(k-p)\, dp\, dk \nonumber \\
&=& \frac{1}{(2\, \pi)^2}\int_{-\infty}^{\infty} e^{ipx} \hat{f}(p) dp \int_{-\infty}^{\infty} e^{ik'x}\,g(k') dk'  \nonumber \\
&=& f(x)\,g(x) \, ,
\end{eqnarray} 
where we substituted with \(k'=k-p\) and used \ref{eqn:ICFT} to write the last line.
Hence, a convolution of the functions \(\hat{f}(k)\) and \(\hat{g}(k)\) in Fourier-space is a product of the respective inverse Fourier transforms \(f(x)\) and \(g(x)\) in real-space.

\subsection{Fourier transform of the sampling operator}
\label{CFT_SAMPLING_OPERATOR}
The sampling function \(\Pi(x)\) is given as an impulse train:
\begin{equation}
\label{eqn:Impulsetrain}
\Pi(x)=\sum_{m=-\infty}^{\infty} \delta^D(x-m\Delta x) \,
\end{equation}
where \(\Delta x\) is the distance between two sampling positions, or the sampling interval.
Applying the Fourier transform \ref{eqn:CFT} to the sampling function \(\Pi(x)\) yields:
\begin{eqnarray}
\label{eqn:CFT_sampling operator_a}
\hat{\Pi}(k) &=&\sum_{m=-\infty}^{\infty} e^{-ik m\Delta x} \,. 
\end{eqnarray} 
As can be easily seen, equation \ref{eqn:CFT_sampling operator_a} has the form of a Fourier series, in fact it is the Fourier series of a Dirac comb except for a constant factor, as we are going to show now.
The Fourier series of a periodic function \(f(z)\) with period \(\Delta z\) is defined as \citep{LangPucker}:
\begin{equation}
\label{eqn:Fourier_series}
f(z)=\sum_{m=-\infty}^{\infty} C_m e^{-i \frac{2\pi}{\Delta z}m\,z} \,
\end{equation}
with the series coefficients \(C_m\) being:
\begin{equation}
\label{eqn:Fourier_series}
C_m=\frac{1}{\Delta z} \int_{0}^{\Delta z} f(z) e^{i\frac{2\pi}{\Delta z}m\,z} dz \, .
\end{equation}
As a Dirac comb \(\Pi(z)\) is periodic with period \(\Delta z\), it can be represented by a Fourier series as:
\begin{equation}
\label{eqn:Fourier_series}
\Pi(z)=\sum_{m=-\infty}^{\infty} \delta^D(z-m\Delta z)=\frac{1}{\Delta z} \,\sum_{m=-\infty}^{\infty} e^{-i\frac{2\pi}{\Delta z}m\,z} \, .
\end{equation}
With this expression we can express \(\hat{\Pi}(k)\) as:
\begin{eqnarray}
\label{eqn:CFT_sampling operator_b}
\hat{\Pi}(k) &=& p_{s}\sum_{m=-\infty}^{\infty} \frac{1}{p_{s}} e^{-ik m\Delta x} \nonumber \\
&=& p_{s} \sum_{m=-\infty}^{\infty} \delta^D\left(k-m\,p_{s}\right) \, ,
\end{eqnarray} 
where we introduced the periodicity length \(p_{s}\) in Fourier-space and the uncertainty principle \(p_{s}\, \Delta x=2\pi \).
Equation \ref{eqn:CFT_sampling operator_b}  shows that the Fourier transform of a Dirac comb is again a Dirac comb. But it is important to note that this proof is only valid under the assumption that the Dirac comb samples at a certain period. As soon as the sampling intervals of the Dirac comb become non-periodic this proof does not apply any longer.

\subsection{Fourier transform pair of the ideal low pass filter}
\label{CFT_ideal_low_pass_filter}
The ideal low pass filter is defined by:
\begin{eqnarray}
\label{eqn:ideal_Low_pass_filter}
\hat{W}(k) = \left \{ \begin{array}{ll}
  1 & \quad \mbox{for $|k|<p_{max}$}\\
  0 & \quad \mbox{for $|k|\geq p_{max}$}\\ \end{array} \right.
\end{eqnarray}
Applying the inverse Fourier transform \ref{eqn:ICFT} we yield:
\begin{eqnarray}
\label{eqn:ideal_Low_pass_filter_real_space}
W(x)&=&\frac{1}{2\pi}\int_{-\infty}^{\infty} \hat{W}(k)e^{ikx} dk \nonumber \\
&=&\frac{1}{2\pi}\int_{-p_{max}}^{p_{max}} e^{ikx} dk \nonumber \\
&=&\frac{1}{i\,2\pi \, x}\left(e^{ip_{max}x}-e^{-ip_{max}x}\right)\nonumber \\
&=& \frac{sin(p_{max}x)}{\pi \,x} \nonumber \\
&=& \frac{p_{max}}{\pi}\, sinc(p_{max}x)
\end{eqnarray}
where we have introduced the sinus cardinal \(sinc(x)=sin(x)/x\).
Hence, the real-space representation of the ideal low pass filter is a sinc function.

\subsection{Fourier transform of the finite sum sampling operator}
\label{Finite_sum_sampling_operator}
The finite sum sampling operator is given as:
\begin{equation}
\label{eqn:APPENDIX_Finite_sampling_operator}
 \Pi_{N}(x) = \sum_{j=-N/2}^{N/2} \delta^D(x-j \Delta x) \, .
\end{equation}
Applying the Fourier transform \ref{eqn:CFT} then yields:
\begin{equation}
\label{eqn:APPENDIX_Finite_sampling_operator_FT}
 \hat{\Pi}_{N}(p) = \sum_{j=-N/2}^{N/2} e^{-\sqrt{-1}\,p\,j\,\Delta x } = \sum_{j=-N/2}^{N/2} \left(e^{-\sqrt{-1}\,p\,\Delta x } \right)^j \, .
\end{equation}
Using the closed form of the geometric series:
\begin{equation}
\label{eqn:Geometric_series_closed}
\sum_{m=L}^{U} a^{m}=\frac{a^L-a^{U+1}}{1-a} \,
\end{equation}
with \(a=e^{-\sqrt{-1}\,p\,\Delta x }\), \(L=-N/2\) and \(U=N/2\) yields:
\begin{eqnarray}
\label{eqn:APPENDIX_Finite_sampling_operator_FT_ASINC}
\hat{\Pi}_{N}(p) & = & \frac{e^{\sqrt{-1}\,p\,\Delta x \frac{N}{2} }-e^{-\sqrt{-1}\,p\,\Delta x(\frac{N}{2}+1) }}{1-e^{-\sqrt{-1}\,p\,\Delta x }} \nonumber \\ & = & \frac{e^{\sqrt{-1}\,p\,\frac{\Delta x}{2} (N+1)}-e^{-\sqrt{-1}\,p\,\frac{\Delta x}{2}(N+1) }} {e^{\sqrt{-1}\,p\,\frac{\Delta x}{2} }-e^{-\sqrt{-1}\,p\,\frac{\Delta x}{2} }} \nonumber \\
& = & \frac{\sin \left (\frac{k\Delta x}{2}(N+1)\right)}{\sin\left(\frac{k\Delta x}{2}\right)} \nonumber \\
& = & \rm{asinc}_{N}(p) \, ,
\end{eqnarray} 
where we have defined the aliased sinc function \(\rm{asinc}_{N}(p)=\sin \left ( (k\Delta x)/2(N+1)\right)/\sin\left((k\Delta x)/2\right)\).

\subsection{Ideal discretization kernel}
\label{Ideal_discretization_kernel}
To discretize the continuous and non-periodic function in real and Fourier space, one has to convolve the signal first with the replication kernel, and then apply a low-pass filter, or vice versa. Thus, the ideal discretization kernel will be a convolution of the ideal low-pass filter, with the replication operator. This can be written as follows:
\begin{equation}
\label{eqn:Geometric_series_closed}
\Psi(x)=\int_{-\infty}^{\infty} dz W(x-z) \Pi_R(z)\, .
\end{equation}
According to the convolution theorem we can express the Fourier transform \(\hat{\Psi}(p)\) as:
\begin{eqnarray}
\label{eqn:ideal_FS_discretisation_operator}
\hat{\Psi}(p)&=& \hat{W}(p) \hat{\Pi}_R(p) \nonumber \\
&=& \frac{2\,\pi}{L}\sum_{j=-\infty}^{\infty} \delta^D\left(p-j\, \Delta p\right)\hat{W}(p)\, .
\end{eqnarray}
Applying an inverse Fourier transform then yields:
\begin{eqnarray}
\label{eqn:ideal_RS_discretisation_operator}
\Psi(x)&=& \frac{1}{L} \sum_{j=-\infty}^{\infty} \int_{-\infty}^{\infty} e^{ipx}    \delta^D\left(p-j\, \Delta p\right)\hat{W}(p) dp\nonumber \\
&=& \frac{1}{L} \sum_{j=-\infty}^{\infty} e^{ij\, \Delta px} \hat{W}(j\, \Delta p)\nonumber \\
&=& \frac{1}{L} \sum_{j=-M/2}^{M/2} \left (e^{i\, \Delta px} \right)^j \, ,
\end{eqnarray}
where we made use of the fact, that the ideal low-pass filter given by \ref{eqn:ideal_Low_pass_filter} vanishes outside the base-band. One can now use the closed form of the geometric series given in equation (\ref{eqn:Geometric_series_closed}) to yield:
\begin{eqnarray}
\label{eqn:ICFT_FT_sampling_operator}
\Psi(x) &=& \frac{1}{L} \frac{e^{-ix\Delta p M/2}-e^{ix\Delta p (M/2+1)}}{1-e^{ix\Delta p}}  \nonumber \\ &=& \frac{1}{L} \frac{e^{ix\frac{\Delta p}{2}}}{e^{ix\frac{\Delta p}{2}}} \frac{e^{-ix\Delta p (M/2+\frac{1}{2})}-e^{ix\Delta k (M/2+\frac{1}{2})}}{e^{-ix\frac{\Delta k}{2}}-e^{ix\frac{\Delta k}{2}}}  \nonumber \\
&=& \frac{1}{L} \frac{sin\left(x\frac{\Delta p}{2} (M+1)\right)} {sin\left(x\frac{\Delta p}{2}\right)} \nonumber \\
&=& \frac{1}{L} asinc_{M}(x) \, .
\end{eqnarray} 
Thus the ideal discretization kernel is an aliased sinc function in real-space.

\subsection{Discrete Fourier transformation}
\label{Discrete_Fourier_transformation}
we define the numerical Fourier transformation as:

\begin{equation}
\label{eqn:FFT}
y_k=\hat{C}\sum_{j=0}^{N-1}x_je^{-2\pi j k \frac{\sqrt{-1}}{N}}
\end{equation}
and the inverse Fourier transform
\begin{equation}
\label{eqn:IFFT}
x_j=C\sum_{k=0}^{N-1}y_k e^{2\pi j k \frac{\sqrt{-1}}{N}} = C\sum_{k=-(\frac{N}{2}-1)}^{\frac{N}{2}}y_k e^{2\pi j k \frac{\sqrt{-1}}{N}}  \, .
\end{equation}
The normalization coefficients \(C\) and \(\hat{C}\) are chosen such that they fulfill the requirement:
\begin{equation}
\label{eqn:Normalisation_condition}
C\hat{C}=\frac{1}{N} \, .
\end{equation}

\subsection{Discrete mode coupling function}
\label{DISC_MODE_COUPLING_FUNC}
Here we proof the equality 
\begin{eqnarray}
\label{eqn:mode_coupling_function_disc_APP}
\hat{U}(\Delta p\, (i - j)) =\hat{C}\,\sum_{k=0}^{N-1}\,e^{ 2 \pi\,\sqrt{-1} k \frac{(j -i)}{N}}=\hat{C}\,N\,\delta^K_{ij} \, ,
\end{eqnarray}
by evaluating the sum.
First we discuss the case \(j\neq i\) where we will consider the sum
\begin{eqnarray}
\label{eqn:Kronecker_delta}
\sum_{k=0}^{N-1}e^{2\pi k \frac{\sqrt{-1}}{N}(j-i)}&=&\sum_{k=0}^{N-1} \cos{\frac{2\pi}{N}k(j-i)}+\sqrt{-1} \sin{\frac{2\pi}{N}k(j-i)} \nonumber \\
&=& 1+\sum_{k=1}^{N-1} \cos{\frac{2\pi}{N}k(j-i)}+\sqrt{-1} \sum_{k=1}^{N-1} \sin{\frac{2\pi}{N}k(j-i)} \nonumber \\
\end{eqnarray}
By making use of the trigonometric sums:
\begin{equation}
\label{eqn:Trig_sum_cos}
\sum_{k=1}^{N-1} \cos(kx)=\frac{\sin\left(\left (N-\frac{1}{2}\right ) x\right)}{2\sin\left(\frac{x}{2}\right)}-\frac{1}{2}
\end{equation}
and
\begin{equation}
\label{eqn:Trig_sum_sin}
\sum_{k=1}^{N-1} \sin(kx)=\frac{\sin(\frac{N-1}{2}x)\sin(\frac{N}{2}x)}{\sin(\frac{x}{2})}
\end{equation}
we yield:
\begin{eqnarray}
\label{eqn:Kronecker_delta_1}
&=& 1+\sum_{k=1}^{N-1} \cos{\frac{2\pi}{N}k(j-i)}+\sqrt{-1} \sum_{k=1}^{N-1} \sin{\frac{2\pi}{N}k(j-i)} \nonumber \\ &=& 1+\frac{\sin\left(\left (N-\frac{1}{2}\right ) \frac{2\pi}{N}(j-i)\right)}{2\sin(\frac{\pi}{N}(j-i))}-\frac{1}{2} \nonumber \\ & &+\sqrt{-1}\frac{\sin\left(\frac{N-1}{N}\pi(j-i)\right)\sin(\pi(j-i))}{sin(\frac{\pi}{N}(j-i))} \nonumber \\
\end{eqnarray}
because of \(\sin(\pi(j-i))=0\) for all \(j\) and \(i\) the last term vanishes, and we yield
\begin{eqnarray}
\label{eqn:Kronecker_delta_calc}
&=& \frac{1}{2}+\frac{\sin\left(2\pi(j-i)-\frac{\pi}{N}(j-i)\right)}{2\sin(\frac{\pi}{N}(j-i))} \, .
\end{eqnarray}
Since the sinus is \(2\, \pi\)-periodic we can write:
\begin{equation}
\label{eqn:Trig_identity_sin_a}
\sin\left(2\pi(j-i)-\frac{\pi}{N}(j-i)\right)=-\sin\left(\frac{\pi}{N}(j-i)\right)
\end{equation}
Using equation \ref{eqn:Trig_identity_sin_a} in equation \ref{eqn:Kronecker_delta_calc} we yield:
\begin{eqnarray}
\label{eqn:Kronecker_delta_calc_a}
&=& \frac{1}{2}-\frac{\sin\left(\frac{\pi}{N}(j-i)\right)}{2\sin(\frac{\pi}{N}(j-i))} \nonumber \\
&=& \frac{1}{2}-\frac{1}{2} = 0 \, .
\end{eqnarray}
Hence, for the case \(j\neq i\) the sum will always be zero.
Now we discuss the case \(j=i\), where the sum yields:
\begin{equation}
\label{eqn:Kronecker_delta_0}
\sum_{k=0}^{N-1}e^{2\pi k \frac{\sqrt{-1}}{N}(j-i)}=\sum_{k=0}^{N-1} 1 = N \, .
\end{equation}
Hence, we end up with the result:
\begin{eqnarray}
\label{eqn:Kronecker_delta}
\sum_{k=0}^{N-1}e^{2\pi k \frac{\sqrt{-1}}{N}(j-i)} = \left \{ \begin{array}{ll}
  N & \quad \mbox{for $i=j$}\\
  0 & \quad \mbox{for $i\neq j$}\\ \end{array} \right.
\end{eqnarray}
Which yields:
\begin{equation}
\label{eqn:Kronecker_delta_definition}
\sum_{k=0}^{N-1}e^{2\pi k \frac{\sqrt{-1}}{N}(j-i)} = N \delta^K_{ij}\, .
\end{equation}
Thus, equation (\ref{eqn:mode_coupling_function_disc_APP}) has been proven.

\bsp

\label{lastpage}

\end{document}